\documentclass[final,5p,times,twocolumn]{elsarticle}

\usepackage{subfiles}
\usepackage{amssymb}
\usepackage{svg}
\usepackage{stfloats}
\usepackage{multirow}
\usepackage[para]{threeparttable}
\usepackage{pifont}
\newcommand{\cmark}{\ding{51}}
\newcommand{\xmark}{\ding{55}}
\usepackage{hyperref}
\usepackage{xtab,afterpage}
\usepackage{soul}

\usepackage{longtable}
\usepackage{balance}
\usepackage{amsmath}

\journal{NEUROCOMPUTING}

\begin{document}

\begin{frontmatter}


\title{A Survey on Deep Learning Based Point-Of-Interest (POI) Recommendations}

\author{Md. Ashraful Islam}
\author{Mir Mahathir Mohammad}
\author{Sarkar Snigdha Sarathi Das}
\author{Mohammed Eunus Ali}

\address{Department of Computer Science and Engineering (CSE)\\ Bangladesh University of Engineering and Technology (BUET)\\Dhaka 1000, Bangladesh}




\begin{abstract}

Location-based Social Networks (LBSNs) enable users to socialize with friends and acquaintances by sharing their check-ins, opinions, photos, and reviews. Huge volume of data generated from LBSNs opens up a new avenue of research that gives birth to a new sub-field of recommendation systems, known as Point-of-Interest (POI) recommendation. A POI recommendation technique essentially exploits users' historical check-ins and other multi-modal information such as POI attributes and friendship network, to recommend the next set of POIs suitable for a user. A plethora of earlier works focused on traditional machine learning techniques by using hand-crafted features from the dataset. With the recent surge of deep learning research, we have witnessed a large variety of POI recommendation works utilizing different deep learning paradigms. These techniques largely vary in problem formulations, proposed techniques, used datasets and features, etc. To the best of our knowledge, this work is the first comprehensive survey of all major deep learning-based POI recommendation works. Our work categorizes and critically analyzes the recent POI recommendation works based on different deep learning paradigms and other relevant features. This review can be considered a cookbook for researchers or practitioners working in the area of POI recommendation.
\end{abstract}





\end{frontmatter}

\section{Introduction}

\label{sec:introduction}

Location-based Social Networks (LBSNs) offer users a unique opportunity to socialize by sharing their check-ins, opinions, photos, and reviews. All these advantages paired with the wide availability of smartphones have dramatically increased the user base to billions-scale in these LBSNs platforms. Consequently, we have witnessed an explosion of rich multimodal spatio-temporal data collected from these platforms. The availability of this huge amount of data opens up new opportunities in Point-of-Interest (POI) recommendation, a vibrant independent sub-area in a recommendation system that has garnered significant attention from both user and business perspectives in recent years. A POI recommendation technique essentially exploits users' historical check-ins and other multimodal information to recommend the next set of POIs suitable for a user. As the size and modality of the data, and the user expectation widely vary, this opportunity of having tons of multimodal data comes up with new challenges enticing the researchers to design novel techniques to better capture mobility patterns and other features (e.g., spatial, social, textual) to improve recommendation performance.

Earlier works in POI recommendation primarily focused on feature engineering and conventional (non-deep learning) machine learning-based methods. Markov Chain based stochastic models have been explored extensively in this regard \cite{rendle2010factorizing,cheng2013you,chen2014nlpmm,zhang2014lore,ye2013s,yue2019multi,wen2019loc2vec,mao2019prme,xiong2020point}. Due to the success of Matrix Factorization (MF \cite{koren2009matrix}) based methods for recommendation systems in other domains, MF methods \cite{lian2014geomf, liu2014exploiting, wang2018exploiting, cheng2013you,su2020fgcrec, pan2019deep, cai2018integrating, rahmani2020joint} have also been studied for better POI recommendation modeling. To achieve better performance than vanilla MF methods, Bayesian Personalized Ranking (BPR \cite{rendle2012bpr}) methods have been employed \cite{he2017category,he2018inferring,cui2019distance2pre,Zhao2016stellar,zhao2017geo,zhao2017geo2,zhao2018stellar,mao2019prme}. Other traditional approaches like support vector machine (SVM) \cite{gao2018personalized}, Collaborative Filtering \cite{liu2018geographical,ye2010location,ye2011exploiting,yang2017bridging,li2019lori,qiao2018socialmix}, Gaussian Modeling \cite{wang2019geography}, Transitive Dissimilarity \cite{baral2019hirecs} have also been exploited in different works for personalized POI recommendation. One major shortcoming of all these approaches lies in feature engineering. Explicit feature engineering requires sufficient domain expertise. The increasing availability of data from other modalities like images, texts, and POI reviews make these feature engineering tasks even more challenging as manually crafting the relationship between these unstructured features is not a trivial task. Consequently, deep learning-based methods replaced most of those traditional techniques in recent years.

Deep learning methods like Convolutional Neural Networks (CNN) or Recurrent Neural Networks (RNN) provide many advantage in terms of automatic feature extraction eliminating the difficulties in handcrafted feature design. Furthermore, deep learning-based methods excel in modeling complex relationships between structured and unstructured data, which let us leverage multimodal data from different domains in POI recommendation.  In the last few years, we have seen an unprecedented rise in the number of works leveraging deep learning in POI recommendation in all major venues (e.g., AAAI, IJCAI, SIGIR, CIKM, WWW, etc.). The use of different deep learning paradigms such as CNN \cite{xu2019ssser, chen2020cem}, RNN \cite{liu2016predicting, yang2020location, zhao2020discovering, wang2020next}, Long Short Term Memory (LSTM) \cite{sun2020go, li2018next, zhang2020interactive, yu2020category}, Gated Recurrent Unit (GRU) \cite{feng2018deepmove, manotumruksa2018contextual, kala2019context}, and self-attention \cite{lian2020geography, guo2020sanst} have greatly boosted the performance of POI recommendation models. On top of that, state-of-the-art techniques from Natural Language Processing (NLP) have also been employed for complex modeling of human mobility in POI recommendation. Some recent works have leveraged graph embedding to enrich models with semantic geospatial information \cite{liu2017learning, christoforidis2019reline, xiong2020dynamic}. The wide varieties of deep learning-based POI recommendation techniques introduced in a short timespan necessitates a comprehensive review of these works (i) to demonstrate how different techniques have been used to handle different features, (ii) to identify the pros and cons of each model, and (iii) to propose a summary guideline for potential gaps and future research opportunities.

\textbf{Previous Surveys on POI recommendation:} 
Several survey papers exist in the literature that prior works on Point-of-Interest (POI) recommendation. In an early work, Bao et al. \cite{bao2015recommendations} reviewed traditional (i.e., non-deep learning) POI recommendation methods. In a later work, Liu et al. \cite{liu2017experimental} did an experimental evaluation of some of the then state-of-the-art traditional POI recommendation models. In another work, Zhao et al. \cite{zhao2016survey} classified POI recommendation models in three taxonomies: influential factor-based, methodology based, and task-based. All those surveys primarily focused on feature engineering-centric (i.e., non-deep learning) models. Later, Zheng et al. \cite{zheng2018survey} did a comprehensive review of location prediction on Twitter dataset, where they also acknowledged the uprise of deep learning centric approaches. In another review of POI recommendation models, Liu et al.\cite{liu2018user} also mentioned a few neural network models. Since then, a large body of works in POI recommendation have been introduced leveraging different deep learning paradigms. In the last few years, researchers utilized RNN, LSTM, CNN, graph neural networks, attention networks in different ways to make use of different features resulting in significant performance uplift. Recently, Wang et al. \cite{wang2020deep} summarized a handful of deep learning-based models in the spatio-temporal domain. However, since this paper was a summary of the whole spatio-temporal domain, few POI recommendation models were discussed in high-level. Another recent review of location prediction models \cite{xu2020survey} mostly discusses non-deep learning models with a very coarse focus on some deep learning models.

Large varieties of recent deep learning-based POI recommendation works largely vary w.r.t. problem formulations, proposed techniques, used datasets, features, etc. There is no unified study to categorically discuss the pros and cons of different deep learning paradigms on POI recommendations. The wide variety of these techniques can easily puzzle someone willing to explore this field of POI recommendation. This survey work fills up the above gaps of existing studies.

\textbf{Contributions of Our Survey:}
In summary, to fill the gaps of existing surveys on POI recommendation techniques and to cover growing number recent papers papers in this domain, we have made the following major contributions in this review paper.

\begin{itemize}
  \item We provide a categorization of all the models of POI recommendation based on their application goals (Section \ref{section:problemdefinition}).
  \item We outline the features of all datasets used in this domain and discuss their strengths as well as their limitations (Section \ref{section:datasetdescription}).
  \item We categorize the POI recommendation models based on different deep learning paradigms and compare their competitive (dis)advantages (Section \ref{section:recommendationmodels}).
  \item We identify different factors (i.e., social influence, sequential effect, etc.) that impact the POI recommendations and provide tabular analysis of each factor that is covered by all the models (Section \ref{section:influentialfactors}).
  \item We present the comparison of all stat-of-of-art techniques based on their performance metrics (Table \ref{paperComapreTable}).
  \item Finally, we identify shortcomings of the existing works and provide comprehensive future recommendations for POI research (Section \ref{section:shortcomings}).
\end{itemize}

\section{Problem Definition}

\label{section:problemdefinition}
Point-of-Interest (POI) recommendation is a class of problems that suggest suitable future POIs for a user, given the historical check-in history of past users and other associated data of an LBSN. Let $U=\{u_1,u_2,... u_{N}\}$ be a set of $N$ LBSN users and $P=\{p_1,p_2,... p_{M}\}$ be a set of $M$ POIs in the LBSN. Users may be linked to each other through a set of connections \textit{\"{U}}$ = \{\langle u_i, u_j \rangle~ | ~u_i, u_j \in U \}$. Each POI $p$ is geo-coded by latitude $x_p$, longitude $y_p$, and a set of attributes $W_p$ representing POI semantics. 
We first define the relevant terms and then formally define the problem as follows:

\subsection*{Definition 1 (Check-in):} A check-in $\rho_{t_i}^u$ indicates the POI checked-in by user $u$ at time $t_i$.

\subsection*{Definition 2 (Check-in List):} Each user $u$ is associated with a list of check-ins $C^u=\{\rho_{t_1}^u, \rho_{t_2}^u, ..., \rho_{t_T}^u\}$, where $\rho_{t_i}^u$ denotes a check-in record of user $u$ at time $t_i$ and $1<=i<=T$.

\subsection*{Definition 3 (Next POI recommendation):} Given a check-in list $C^u$ of a user $u$, next POI recommendation refers to the prediction of the next POI at time $t_{T+1}$.

\subsection*{Definition 4 (Sequence of POI Recommendation):} Given a check-in list $C^u$ of a user $u$, sequence of POI recommendation will recommend the next $n$ POIs which is from $t_{T+1}$ to $t_{T+n}$. 

\subsection*{Definition 5 (Missing POI Check-in Identification):} Given the check-in list, $C^u=\{\rho_{t_1}^u, \rho_{t_2}^u, ..., \rho_{t_{m-1}}^u, \rho_{t_m}^u, \rho_{t_{m+1}}^u, ..., \rho_{t_T}^u\}$ of an user $u$ where, the check-in $\rho_{t_m}^u$ is missing then missing POI check-in identification will identify the POI at time $t_m$. \\


\section{Network Architecture Preliminaries}

In this section, we present the preliminary overview of different deep neural network paradigms that include Feed-Forward Network, Convolutional Neural Network, Recurrent Neural Network, Long-Short Term Memory, Gated Recurrent Unit, Attention Mechanisms, and Generative Adversarial Network.

\subsection{Feed-Forward Network}
        \begin{figure}[ht]
        \begin{center}
        \includegraphics[width=0.4\textwidth]{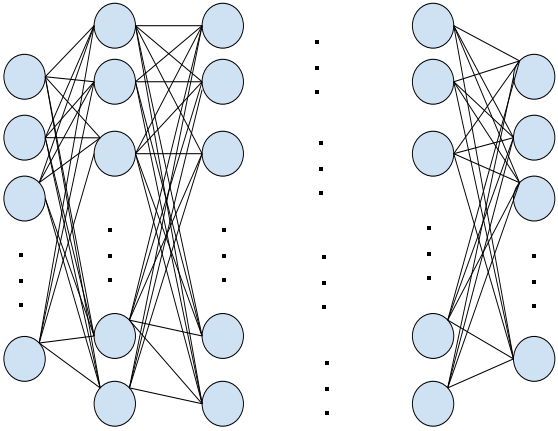}
        \caption{Feed-Forward Network}
        
        \label{fig:DNNFigure}
        \end{center}
        \end{figure}
        
        Feed-Forward Networks (Figure~\ref{fig:DNNFigure}) are the most basic form of neural networks. Neural nodes are stacked up in layers where every node from a layer is connected to all the nodes in the next layer. The weighted connections combine the features of one layer and pass them to the subsequent layer through a nonlinear function (e.g. ReLU, Sigmoid, tanh, etc.). Stacking up layers of neurons dramatically increases the expressiveness of the network.
        
        Although feed-forward networks can capture highly complex relationships within features, their overly high representational power usually causes overfitting training data resulting in poor generalization. Furthermore, as the number of layers increases, the sizes of the models dramatically increase making them harder to train and deploy. Most importantly, feed-forward networks have no explicit spatial and sequential feature handling capability, which limits their usage in spatio-temporal models.
        
    \subsection{Convolutional Neural Network}
\begin{figure}[ht]
        \begin{center}
        \includegraphics[width=0.45\textwidth,height=0.12\textheight]{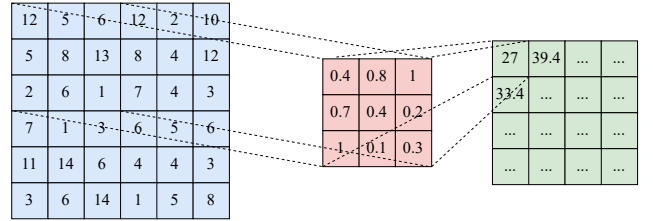}
        \caption{CNN Layer Workflow}
        
        \label{fig:cnn_pic}
        \end{center}
        \end{figure}

        Convolutional Neural Networks (CNNs) are particularly suited for capturing spatial features from a given input. In CNN, convolutional filters, and pooling layers are systematically used to hierarchically process inputs. Subsets of the inputs are gradually channeled through convolutional filters, and pooling layers are used to scale down the transformed features. This process helps the CNN to gain spatial awareness while keeping the number of parameters significantly lower than feed-forward networks. To extract the spatial patterns in spatiotemporal data, CNNs are thus proven to be highly useful. Figure~\ref{fig:cnn_pic} shows the workflow of a CNN layer with filter size of 3x3.
        
\subsection{Recurrent Neural Network}
\begin{figure}[ht]
        \begin{center}
        \includegraphics[width=0.4\textwidth]{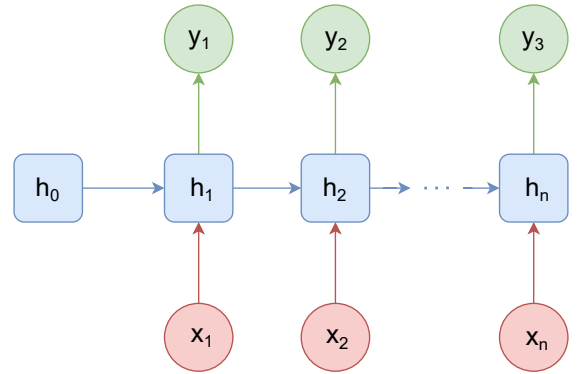}
        \caption{RNN Model}
        \label{fig:RNNFigure}
        \end{center}
        \end{figure}

        Feed-forward networks or Convolutional Neural Networks do not consider any sequential or temporal dependency within the inputs. Recurrent Neural Network (RNN) solves this problem by taking a sequence of inputs and then learning the sequential pattern of the input sequence by using hidden states. We show the basic diagram of RNN in Figure \ref{fig:RNNFigure}. In this figure $x_i, y_i$ and $h_i$ represent the input, output, and hidden state, respectively. We can see that besides input each RNN block uses a hidden state to produce output. Actually, the hidden states capture the context information of the input sequence which means capture the sequential pattern.
    \subsection{Long-Short Term Memory}
\begin{figure}[ht]
        \begin{center}
        \includegraphics[width=0.45\textwidth]{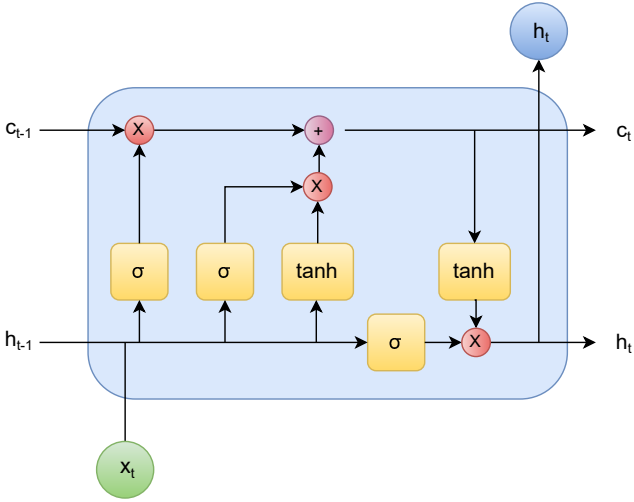}
        \caption{Basic LSTM Cell}
        \label{fig:LSTMFigure}
        \end{center}
        \end{figure}

        RNN suffers from exploding Gradients and vanishing Gradients problems, as a result, can not capture long-term preferences. To solve the problems Long-Short Term Memory (LSTM) \cite{lstmarticle} is proposed. LSTM uses a gate mechanism and can capable of capturing long-term preferences. We can see an LSTM cell in Figure \ref{fig:LSTMFigure}. In addition to hidden state $h_t$ which is used in RNN, every LSTM block has a cell state $c_t$. Also, the flow of the information among consecutive LSTM cells are controlled via three gates: (1) input gate, (2) forget gate, and (3) output gate. 
        
    \subsection{Gated Recurrent Unit}
        LSTM resolves the problems of RNN but it has three gates so the training of an LSTM based model is slower and requires a large amount of training data. For solving these issues Gated Recurrent Unit (GRU) \cite{cho2014learning} is proposed. It uses only two gates, i.e., reset and forget gates. Thus, the GRU based model can be trained faster and performs better than LSTM when there is less training data.  
    
    \subsection{Attention Mechanism}
        Sequence models like RNNs or LSTMs process inputs by logical order of sequence. However, this scheme tends to lose features in longer sequences resulting in poor model performance. Attention mechanism \cite{bahdanau2014neural} largely solves this shortcoming by mimicking a humanlike focus in salient input regions. Humans are prone to giving higher attention to key parts of the input, which in turn helps to break down a complex input into simpler parts that can easily be processed. While Seq2Seq models~\cite{sutskever2014sequence} have taken the advantage of this attention mechanism to improve performance, recent advancements in attention mechanism have introduced self-attention mechanism \cite{vaswani2017attention} that improves performance as well as allows parallel processing of inputs making them lucrative for various applications. The key idea here is that inputs are mapped to query, key, and value vectors. The outputs are calculated by taking the weighted sum of the value vectors where weights are determined by a function of query and key values. This technique has been highly effective in many areas of NLP research, which shows its potential in other domains involving sequential data.
        
        \subsection{Generative Adversarial Network}
    
\begin{figure}[ht]
        \begin{center}
        \includegraphics[width=0.45\textwidth]{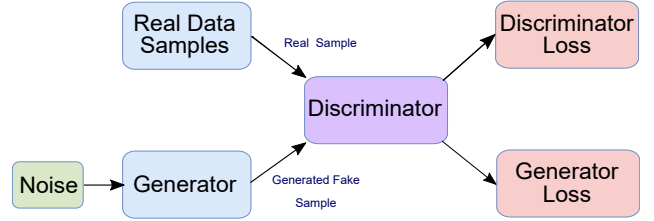}
        \caption{Generative Adversarial Network}
        \label{fig:GANFigure}
        \end{center}
        \end{figure}

        Generative adversarial networks (GANs) \cite{goodfellow2014generative} are a special form of generative machine learning framework where two different networks compete against each other with different goals. One of them is called ``Generator" and the other is called ``Discriminator". The generator network generates candidates whereas the discriminator network tries to assess those candidates. The generator tries to ``fool" the discriminator by creating novel sample candidates whereas the discriminator tries to distinguish those samples from the true data distribution. A generic GAN is shown in Figure \ref{fig:GANFigure}. In recent years, GANs have shown impressive performance in image synthesis, video game resolution upscaling, art generation, and so on. 

\section{Dataset Description}

\label{section:datasetdescription}

Prior works in Point-of-Interest recommendation have used check-in data collected from a wide range of LBSNs that include Foursquare, Gowalla, Yelp, Twitter, Facebook, Brightkite, Instagram, WeChat, and Baidu Map. Most of these datasets consist of tabular data that records the user-POI and the user-user relationship in LBSNs. User-POI data typically contains user check-in information including timestamps, location, and semantic features. Here POI Semantic features include the categories of the POIs and tags included in user LBSN posts, creation date of the POIs, geolocations (latitude, longitude), check-in counts, number of users checked-in, radius, etc. 
On the other hand, user semantic features contain the number of posts, friends, check-ins, etc. In order to keep the social influence in context, datasets like Foursquare, Gowalla, Weeplace also contain user-user relationship as a many-to-many schema, where each user is connected to all his friends. Because of all these data and a huge number of check-ins, some of these datasets become exceedingly large. Consequently, most of the prior works focused on a specific region or a country to keep the size tractable. A brief discussion on these LBSN datasets is given below:

\textbf{Foursquare:}
Founded in 2009, Foursquare\footnote{https://foursquare.com/} has worked with world-wide collection and distribution of location data to facilitate technological corporations and brands. Most of the POI recommendation models discussed in this review use the datasets of Foursquare from a time range of 2010 to 2014. The datasets contain check-in data collected mostly from the USA and Tokyo. This dataset also contains the list of all friends of each user in the LBSN.

\textbf{Gowalla:}
Gowalla is a location-based social media platform dedicated to location check-ins. The platform was founded in 2007 and acquired by Facebook in 2012. Gowalla was primarily a mobile application that allowed users to check into locations that they visited using their mobile devices. The datasets from the functioning period of Gowalla were available via the Gowalla API and currently, there are no official distributors for the datasets. Gowalla is the second most used datasets in the POI recommendation models discussed in this paper. Most prior works discussed have used check-in data from February 2009 to October 2010. Like Foursquare, the Gowalla dataset also contains the list of friends of every user in the dataset. Besides, a detailed description of each POI and user profiles are also available in this dataset.

\textbf{Brightkite:}
Brightkite was a location-based social media network that launched in 2007 and got dissolved in 2012. The platform provided the ability to check-in through text messaging or a mobile application after visiting a location. Datasets of Brightkite is no longer officially available but still can be accessed from various research archives around the globe.

\textbf{Yelp:}
Yelp \footnote{https://www.yelp.com/} is another popular LBSN platform. As users tend to check-in different business locations, Yelp provides reviews and ratings from customers who shared their own experiences both for personal and research purposes. Yelp was founded in 2004 and is still operational as a reviewing company for business establishments. POI recommendation models are greatly benefited from the textual reviews of Yelp because reviews provide semantic information of POIs.

\textbf{Weeplaces:}
WeePlaces is a service that visually maps users' check-ins on location-based services. Weeplaces has been integrated with Gowalla and Facebook, giving users the ability to visualize where users have announced their locations to friends across Foursquare, Gowalla, and Facebook Places. Like Foursquare and Gowalla dataset, Weeplace dataset also contains the list of friends connected to a user which can be leveraged for capturing social influence in POI recommendation.

\textbf{Instagram:}
Instagram \footnote{https://www.instagram.com} is a social media platform initially released in 2010 and currently owned by Facebook. Instagram allows users to post and share photos and videos online. Users can browse other users' content by tags and locations. Instagram dataset primarily contains user check-in data which has been used in some recent works on POI recommendation.

\textbf{Twitter:}
Twitter \footnote{https://twitter.com} is a highly popular social media platform where a user can post, see, and share short messages known as 'tweets'. Twitter was founded in 2006 and is currently the most popular micro-blogging service around the world. The check-in functionality of Twitter enables users to record visits to locations. The datasets are available through Twitter public API. 

\textbf{Other datasets:}
Some other used datasets are collected from WeChat\footnote{https://www.wechat.com/}, Baidu Maps \footnote{https://map.baidu.com/}, Facebook\footnote{https://facebook.com}. Functionalities of WeChat include texting, voice messaging, video conferencing, and location sharing. Baidu Maps provide street maps and views, satellite views of terrains, and route planners for traveling. Facebook is currently the largest social media site where users can post text, photos, and multimedia to share information about themselves. Posts can include check-in information about a visited place and thus datasets of Facebook are officially available for POI prediction research.

For detailed information on some of the most used datasets used in POI recommendation, see Table \ref{datasetsummary}. To see the statistics of the datasets associated with each of the discussed papers, see Table \ref{datasetModel}.

\begin{table*}[!t]
\small
\caption{Summary of Datasets}
\label{datasetsummary}
\centering
\begin{tabular}{ |c|p{65mm}|p{85mm}|}
    \hline
    \textbf{Dataset name} &
    \textbf{Column names} & 
    \textbf{Description of the table} \\
    \hline
    \multirow{2}{*}{Foursquare} &
    userID, Time(GMT), VenueId, VenueName, VenueLocation, VenueCategory	 & 
    Describes the user-POI relationship. Contains all user-checkins with date-time of checkin; name, location and type of POI\\
    \cline{2-3}
    & 
    userID, friendID	 & 
    Describes the user-user relationship. Contains the list of all friends of each user in the LBSN\\
    \hline
    \multirow{4}{*}{Gowalla} &
    userId, timestamp, latitude, longitude &
    Describes the user-POI relationship. Contains all checkin information of each user such as location and time of the visit\\
    \cline{2-3}
    &
    userId, friendId &
    Describes the user-user relationship. Contains the list of all friends of each user in the LBSN\\
    \cline{2-3}
    & 
    id, name, created\_at, lng, lat, photos\_count, checkins\_count, users\_count, radius\_meters, highlights\_count, items\_count, max\_items\_count, spot\_categories, city\_state &
    Description of each POI of the LBSN including the counts of user, checkins, photos, items and highlights. The name, city, date-time, category and radius on the map of each POI are also included.\\
    \cline{2-3}
    & 
    id, bookmarked\_spots\_count, challenge\_pin\_count, country\_pin\_count, highlights\_count, items\_count, photos\_count, pins\_count, province\_pin\_count, region\_pin\_count, state\_pin\_count, trips\_count, friends\_count, stamps\_count, checkin\_num, places\_num &
    Details of the profile of each user of the LBSN\\
    \hline
    
    \multirow{2}{*}{Brightkite} &
    user, check in time, latitude, longitude, location id &
    Describes the user-POI relationship. Each row contains time and location information of check-ins made by one user\\
    \cline{2-3}
    &
    userid1, userid2 &
    Describes the user-user relationship. Friendship network of Brightkite users are described in this table.\\
    \hline
    
    \multirow{6}{*}{Yelp} &
    business\_id, name, address, city, state, postal\_code, latitude, longitude, stars, review\_count, is\_open, attributes, categories, hours &
    Contains POI business data including location data, attributes, and categories. Attributes include whether the restaurants accepts takeouts and has business parkings\\
    \cline{2-3}
    &
    review\_id, user\_id, business\_id, stars, date, text, useful, funny, cool &
    Contains full review text data including the user\_id that wrote the review and the business\_id the review is written for.\\
    \cline{2-3}
    &
    user\_id, name, review\_count, yelping\_since, friends, useful, funny, cool, fans, elite, average\_stars, compliment\_hot, compliment\_more, compliment\_profile, compliment\_cute, compliment\_list, compliment\_note, compliment\_plain, compliment\_cool, compliment\_funny, compliment\_writer, compliment\_photos &
    User data including the user's friend mapping and all the metadata associated with the user.\\
    \cline{2-3}
    &
    business\_id, date &
    Checkins on a POI of all the users\\
    \cline{2-3}
    &
    text, date, compliment\_count, business\_id, user\_id &
    Tips written by a user on a POI business location. Tips are shorter than reviews and tend to convey quick suggestions.\\
    \cline{2-3}
    &
    photo\_id, business\_id, caption, label &
    Contains photo data including the caption and classification\\
    \hline
    
    \multirow{2}{*}{Weeplace} &
    userid, placeid, datetime, lat, lon, city, category &
    Describes the user-POI relationship. Each row contains a check-in information of a user, date and time of the visit. The row also contains the location, category, subcategory and name of the city of the POI.\\
    \cline{2-3}
    &
    userid1, userid2 &
    Describes the user-user relationship. Contains the list of all friends of each user in the LBSN\\
    \hline
    
    \multirow{1}{*}{Instagram} &
    user\_id, latitude, longitude, timestamp &
    Details of the user-POI relationship. Each row contains a check-in information of a user, date and time of the visit and the location of the POI\\
    \hline
    
    \multirow{1}{*}{Twitter} &
    userID, tweetID, latitude, longitude, time, placeID, contentInfo &
    Contains information of a tweet having a check-in. Each row represents a tweet, a user, location and identifier of the POI and tags associated with the POI\\
    \hline

\end{tabular}
\end{table*}

\section{POI Recommendation Models}

\label{section:recommendationmodels}
Thanks to the astounding growth of the user base in LBSNs, the amount of check-in data collected from these platforms have increased rapidly in recent years. This large volume of data has fueled the adaptation of deep learning techniques in the field of POI recommendation. While earlier works used conventional machine learning models, recent deep learning-based models have mostly replaced them due to the significantly higher performance with abundant potential to further improve the performance. Thus, in this survey, we mainly focus on Deep Neural Network (DNN) based POI recommendations. We categorize all of the proposed models used thus far into six major categories. They are RNN based models (Section \ref{BasicRNNmodel}), LSTM models (Section \ref{LSTMmodel}), GRU models (Section \ref{GRUmodel}), Graph Embedding models (Section \ref{GRAPHmodel}) GAN models (Section \ref{GANmodel}), and other models (Section \ref{OTHERmodel}). These categories are described in the following subsections. We briefly highlight these models in Table \ref{paperModel}, and the dataset used in each model are depicted in Table \ref{datasetModel}. A concise summary of these models, evaluation metrics, and their performance across different datasets are given in Table \ref{paperComapreTable}.

    \subsection{RNN based models}
    \label{BasicRNNmodel}
        Recurrent Neural Networks (RNN) are renowned for their high effectiveness in NLP problems. As POI recommendation problems show similar properties that resemble NLP tasks, many recent POI recommendation models use RNN as their base architecture. In this subsection, we discuss the basic RNN based POI recommendation models that can map a POI-sequence to another POI-sequence (successive POI recommendation), or only one POI (next POI recommendation). 
        
        Liu et al. \cite{liu2016predicting} proposed a model, called \textbf{Spatial-Temporal Recurrent Neural Networks (ST-RNN)}, for POI recommendations. The ST-RNN model extends the RNN model for capturing spatial and temporal effects. This model adds a time-specific and distance-specific transition matrix for capturing temporal cyclic effect and geographical influence, respectively. The model also applies linear interpolation for the training of the transition matrix.
        
        In another recent work, Yang et al.\cite{yang2020location} proposed a model called \textbf{Flashback} in which they use Basic RNN. The model uses sparse user mobility data by focusing on rich spatio-temporal contexts and doing flashbacks on hidden states in RNNs. Furthermore, the model uses the weighted average of historical hidden states for better capturing the spatio-temporal effects. Additionally, the paper also uses user embedding for considering user preferences.

        Zhao et al.\cite{zhao2020discovering} proposed \textbf{Adaptive Sequence Partitioner with Power-law Attention (ASPPA)}  model to learn the latent structures of the check-in sequences. The idea is a blend of Adaptive Sequence Partitioner (ASP) for texts by Griffiths et al.\cite{griffiths2004hierarchical} and the stacked RNN architecture of El et al.\cite{el1996hierarchical}. The paper aims to automatically detect and identify each semantic subsequence of POIs and discover their sequential patterns. The model is designed to be a stacked RNN and it adopts a binary boundary detector to control the pattern of cell update. This model uses the Power-law Attention (PA) mechanism to integrate spatial and temporal contexts of each check-in into the model. The output layer consists of two fully connected layers and a drop-out layer. 
        
        Since most of the POI recommendations are designed upon a cloud-based paradigm, there are many disadvantages including privacy concerns. To alleviate these problems, Wang et al.\cite{wang2020next} proposed \textbf{Light Location Recommender System (LLRec)}. Here the authors introduced teacher and student models. The teacher model is deployed on the server whereas the student model is deployed on the user mobile devices. The student model will fetch the pre-trained model from the server. And a user can get the next POI recommendation by using the pre-trained model without sending data to the teacher model. However, the storage and the computational capabilities in mobile devices are very limited. Consequently, the student model must be efficient, lightweight, and fast. For getting a lightweight and fast model, LLRec uses FastGRNN \cite{kusupati2018fastgrnn}. The parameters of this model are further compressed by using the tensor-train format \cite{ma2019tensorized}. Furthermore, the knowledge distillation framework \cite{hinton2015distilling} is used to improve prediction quality with very limited data. Time-specific and distance-specific transition matrix in vanilla FastGRNN is used for capturing spatio-temporal correlations between two adjacent check-ins. 
        On the other hand, a powerful and computationally intensive teacher model can be deployed in the cloud. Besides, The teacher model uses attention mechanism to learn user preferences and converts textual content of POIs into low dimensional embeddings via Word2Vec \cite{mikolov2013efficient}. Besides, Huang et al.\cite{huang2020deep} and Liao et al.\cite{liao2018multi} also extensively used the RNN architecture in their POI recommendation model.
        
    \subsection{LSTM models}
    \label{LSTMmodel}
        Since RNN models cannot capture long term dependencies, Long-Short Term Memory (LSTM) \cite{lstmarticle} has been extensively used in the recent works in POI recommendations. Most of the models use the basic unaltered LSTM for their predictions. Some of the models modify the basic LSTM model or use the bidirectional variant for the better capturing of POI domain-specific factors. To further improve long term dependency modeling, attention mechanism have
        also been used alongside LSTM. A detailed discussion of different kinds of LSTM based POI recommendation models are given below:
        
        \subsubsection{Basic LSTM}
            Most of the models use an encoder-decoder scheme where the encoder usually collects the information from users' check-in list and other attributes; and using this information, the decoder predicts the next POI or a sequence of POI.\\ 
            
            We start our discussion in the Basic LSTM model with the model proposed by Kong et al. \cite{kong2018hst} called \textbf{Hierarchical Spatial-Temporal Long-Short Term Memory (HST-LSTM)}. HST-LSTM uses a hierarchical model to encode the periodicity of people's movement. This model captures users’ historical visiting sequences in an encoder and decoder manner which improves the performance of POI recommendation.
            
            Li et al.\cite{li2018next} proposed \textbf{Temporal and Multi-level Context Attention (TMCA)} that uses LSTM based encoder-decoder network and three types of attention: multi-level context attention (micro, macro) and temporal attention. The paper introduces two attention mechanisms to select relevant historical and contextual factors. The model also uses embedding to incorporate heterogeneous contextual factors in a unified way.
            
            For capturing both long-term and short-term preference Wu et al.\cite{wu2019long} proposed \textbf{Long- and short-term preference learning model (LSPL)}. LSPL has two modules i.e., (1) the long-term module consists of embedding layer and attention layer and captures long-term preference of a user by learning contextual features of POIs; (2) the short-term module uses two separate LSTM modules: one for location level and other for category level and captures sequential behavior of a user. Then all of them are combined to predict the next POI.
             
            The authors of LSPL \cite{wu2019long} extend their work by proposing \textbf{Personalized Long- and Short-term Preference Learning (PLSPL)} \cite{wu2020personalized} model. PLSPL adds a user-based linear combination unit with their existing LSPL model which captures the personalized preferences for different users by learning the personalized weights over long- and short-term modules.
            
            Based on embedding, LSTM, an attention mechanism, Doan et al.\cite{doan2019attentive} proposed \textbf{Attentive Spatio TEmporal Neural (ASTEN)} model. ASTEN embeds the POIs and represents a check-in efficiently. LSTM uses the POI and check-in representation and along with the attention mechanism which captures the sequential, temporal, and geographical influence. The model addresses the noise in user-trajectory data by attention mechanism.
            
            Sun et al.\cite{sun2020go} proposed \textbf{Long and Short-Term Preference Modeling (LSTPM)} using basic LSTM models. LSTPM divides all check-ins into several trajectories. The model actually develops three modules including the long-term preference modeling, the short-term preference modeling, and the prediction module. Long-term preference modeling uses all the trajectories, short-term preference modeling uses the last trajectory, and combinedly predicts the next POI. Another important aspect of the next POI recommendation is that the next POI does not depend only on the recent check-in; however, it can depend on any earlier check-in. But RNN/LSTM based approaches have a drawback of being unable to model the relations between two nonconsecutive POIs. For capturing this affect the model uses a \textbf{geo-dilated LSTM} scheme along with basic LSTM in short-term preference modeling.
            
            Zhang et al.\cite{zhang2020interactive} proposed \textbf{Interactive multi-task learning (iMTL)}, which uses a two-channel encoder and a task-specific decoder. The two-channel encoder (temporal-aware activity and spatial-aware location preference encoders) aims to capture the sequential correlations of activities and location preferences. The representations encoded by the LSTM are utilized in the task-specific decoder to interactively perform the prediction tasks. A novel contribution of this paper is that they focus on the collective POIs. Suppose, $l_1$ is a building and each building contains many individual apartments i.e., $l_2,l_3,l_4,l_5$, then the model denotes $l_1$ as collective POI and $l_2,l_3,l_4,l_5$ as individual POIs. This paper proposes a fuzzy characterization strategy for better prediction of individual POI from a collective POI.

            Most of the POI recommendation models only predict the location of POI ignoring the timestamp but Yu et al.\cite{yu2020category} proposed \textbf{Category-aware Deep Model (CatDM)} which predicts POIs that will be visited by users in the next 24 hours. CatDM contains (1) metric embedding that learns the latent features of a user, POI, POI category, and time; (2) first deep encoder for capturing user preferences in POI categories; (3) two filters for reducing search space to generate candidates; (4) another deep encoder for user preferences in POIs, and (5) a module for ranking the candidate set. For ranking candidate set this model considers four correlations simultaneously i.e., the correlation between user and POI, the correlation between the user and POI category, the correlation between POI and temporal influence, and the correlation between POI and user’s current preferences. The model uses POI categories and geographical influence for overcoming data sparsity. The model also uses an attention mechanism for getting better results.
            
            Let us assume that $l_1, l_2, l_3$ are three locations and $l_2$ is nearby to $l_1$. Now, if a check-in from $l_2$ to $l_3$ presents in a user check-in list then after visiting $l_1$ it will most likely to visit $l_3$. This type of situation is called transition regularity. Most of the existing POI recommendation methods capture sequential regularity only. On the other hand, Gua et al. \cite{guo2020attentional} proposed \textbf{Attentional Recurrent Neural Network (ARNN)} model which captures both sequential and transition regularities for resolving sparsity problem. ARNN consists of several layers i.e., (1) Neighbor discovery layer: neighbors are extracted from heterogeneous data by using knowledge graph (KG) and meta-path; (2) Embedding Layer: transforms the sparse features of check-in sequence into dense representation and learns the spatio-temporal features, semantic context by using multi-modal embedding layer; (3) Attention layer: calculates the similarity between the current location and each neighbor and capture the transition regularities of the neighbors; (4) Recurrent layer: captures higher-order sequential regularity by using LSTM. Besides these model, Wang et al.\cite{wang2020modeling}, Huang et al.\cite{huang2019attention} and Li et al.\cite{li2019multi} uses LSTM and attention layers for POI prediction. Yang et al.\cite{yang2019next} uses LSTM and embedding layers for the next POI recommendation. Another model called \textbf{SLSTM} \cite{zhao2018personalized} uses stacked LSTM and embedding layers for sequential check-in prediction.
            
        \subsubsection{Bi-LSTM}
            Vanilla LSTM processes input only in one direction sequentially. While it helps the models get sequential information from the previous inputs, information from later parts of the input cannot be captured. Bi-directional LSTM (Bi-LSTM) solves this problem by considering both directions of inputs. Some recent POI recommendation models also use Bi-LSTM to capture sequence features from both directions to achieve better performance. \\
            
            
            From the model GeoIE \cite{wang2018exploiting} we find that POI has two propensities i.e., (1) Geo-influence: directs its visitors to other POIs and (2) Geo-susceptibility: the receipt of visitors from other POIs. By considering the above-mentioned properties, Liu et al. \cite{liu2020exploiting} proposed a model called \textbf{Geographical-Temporal Awareness Hierarchical Attention Network (GT-HAN)}. For better capturing the great variation in geographical co-influence across POIs GT-HAN uses three factors i.e., the geo-influence of POIs, the geo-susceptibility of POIs, and the distance between POIs. The main part of GT-HAN is an embedding layer, a geographical-temporal attention network layer, and a context-specific co-attention network layer. The embedding layer captures geo-influence, geo-susceptibility, and semantic effects. The geographical-temporal part explores the geographical relations between POIs and the temporal dependency of a check-in list and uses the Bi-LSTM model to capture the sequence dependence of a user’s check-in list. The context-specific co-attention network captures dynamic user preferences. GT-HAN \cite{liu2020exploiting} is actually the improved version of a previous model that is also known as GT-HAN \cite{liu2019geographical} and proposed by the same authors.
            
            Liu et al. \cite{liu2020time} proposed \textbf{time-aware Location Prediction (t-LocPred)} model. This model has two basic parts i.e., ConvAoI and mem-attLSTM. ConvAoI uses the CNN layer and ConvLSTM layer\cite{liu2014exploiting} to find the correlations among adjacent AoIs (Area-of-Interest) and time periods within a day and a week respectively. So, CNN and ConvLSTM work as short-term and long-term coarse-grained spatial-temporal modeling respectively. On the other hand, mem-attLSTM captures complex long-term correlation using a spatial-aware memory augmented LSTM and time-aware attention mechanism. So, mem-attLSTM works as a fine-grained filter that selects the most likely POIs a user will visit.
            
            Chang et al.\cite{chang2020content} proposed \textbf{Content-aware successive POI recommendation (CAPRE)} which is a complete POI recommendation model that uses user-generated textual content. CAPRE has four modules: (1) input layer: takes check-in history as input; (2) content encoder layer: uses character-level CNN (Convolutional Neural Network), multi-head attention mechanism, and POI embedding for capturing various perspectives of user interests about POIs; (3) user behavior pattern: captures content-aware and geographical user behavior patterns using Bi-LSTM; (4) Output layer:  multi-layer perceptron (MLP) to capture users’ general preferences for POIs.
        \subsubsection{Modified LSTM}
            Some models modify the basic LSTM\cite{lstmarticle} model for enhanced POI prediction. The underlying idea is to better capture user short-term and long-term preferences by modifying the basic LSTM. 
            
            Zhao et al.\cite{zhao2019stgn} proposed \textbf{Spatio-Temporal Gated Network (STGN)} that modifies the basic LSTM \cite{lstmarticle} to capture short-term and long-term preference easily. This model adds four new gates i.e., two for long-term preferences and the other two for short-term preferences. This model also adds a new cell state. So, in the proposed model, there is one cell state for short-term preferences and one cell state for long-term preferences. STGN model is further improved by using coupled input and forget gates called STCGN (Spatio-Temporal Coupled Gated Network). STCGN reduces the number of parameters and thus this model can be trained easily and improves efficiency.
        \subsubsection{Self-Attention}
            Following the success of self-attention in language modeling, state-of-the-art POI recommendation models have leveraged this powerful approach to achieve best in class performance. Among them, Lian et al.\cite{lian2020geography} proposed \textbf{Geography-aware sequential recommender based on the Self-Attention Network (GeoSAN)} uses a geography-aware self-attention network and geography encoder. The attention network consists of an embedding layer, a self-attention encoder, a target-aware attention decoder, and a matching function. The geography encoder uses map gridding and GPS mapping for encoding GPS location as quad keys. To address the sparsity challenge, the paper proposes a weighted binary cross-entropy loss function based on importance sampling, so that informative negative samples are more weighted. 
        
            In another work, Guo et al. \cite{guo2020sanst} proposed a model - \textbf{self-attentive networks along with spatial and temporal pattern learning for next POI recommendation (SANST)}. SANST is actually a modification of the model SASRec \cite{kang2018self} which uses a two-layer \textit{transformer network} \cite{vaswani2017attention}. SASRec does not capture spatial and temporal patterns. So, for capturing spatial patterns, SANST updates the embedding of check-ins by adding the location of the checked-in POI. Since the co-ordinate representation of a POI location is sparse, the model discretizes the whole data space with a grid such that POI locations are represented by the grid cell IDs. Grid Cell IDs are learned using \textit{GeoHash} encoding and Bi-LSTM network. For capturing temporal effect SANST adds a parameter in the self-attention network which is computed using the time difference between two check-ins.

    \subsection{GRU}
    \label{GRUmodel}
        RNN are great for analyzing time series data and LSTM helps in capturing short and long-term effects of visiting POIs. However, both RNN and LSTMs tend to suffer from cold start problems. Most RNN models rely on the last hidden layer which limits the learning of user information from the hidden layers. To solve such cases, a modified form of LSTM- GRU\cite{cho2014learning} has been introduced in many POI recommendation models. The GRU has fewer parameters to learn comparing to LSTM cells but it has additional gates such as forget gate to compensate for the problems mentioned above alongside solving exploding and vanishing gradient problems. \\
        
        A popular model called \textbf{DeepMove} is proposed by Feng et al. \cite{feng2018deepmove} . DeepMove actually predicts human mobility which is very similar to our POI recommendation. DeepMove has two modules i.e., (1) Multi-modal Recurrent Prediction Framework: extract features by jointly embedding spatiotemporal and personal feature into a dense representation, which is then fed into GRU unit to model long-range and complex dependencies in a trajectory sequence; (2) Historical Attention Module: captures multi-level periodicity of human mobility.
        
        Most of the model considers identical impact from different types of contexts on the users' preferences. But their impacts are not identical. To solve this problem Manotumruksa et al. \cite{manotumruksa2018contextual} proposed \textbf{Contextual Attention Recurrent Architecture (CARA)} model. CARA has four layers i.e., input layer, embedding, recurrent layer, output layer. In the recurrent layer this model uses GRU. For capturing different contextual impact on the users' preferences this model uses two types of gating mechanisms i.e., (1) Contextual Attention Gate (CAG): controls the influence of ordinary and transition contexts on the users’ dynamic preferences and (2) Time- and Spatial-based Gate (TSG): considers the time intervals and geographical distances between successive check-ins to control the influence of the hidden state of previous GRU units.
        
        Kala et al. proposed \cite{kala2019context} \textbf{Multi-GRU (MGRU)} which modifies the basic GRU unit by adding two additional gates for a better recommendation. The first added gate is Dynamic Contextual-Attention-Gate (DCAG-$\alpha$) which captures the effect of dynamic contexts like - time of the day, companion, user’s mood, etc. The other gate is Transition-Contextual-Attention-Gate (TCAG-$\beta$) captures the effect of transition contexts like - time interval and geographical distance from past POI to future POI. MGRU has three layers i.e., (1) input layer: pre-process and embeds check-in sequence; (2) recurrent layer: captures sequential patterns using MGRU; (3) output layer: recommend the next POI.
    
    \subsection{Graph Embedding}
    \label{GRAPHmodel}
        Some recent techniques leverage the potentials of Graph Embeddings (GE) that learn low-dimensional key features of the dataspace modeled as different forms of graphs such as POI-POI, user-POI, and POI-time.
        
        Xie et al.\cite{xie2016learning} proposed \textbf{GE} that uses graph embedding for recommending next POI. GE jointly captures the sequential effect, geographical influence, temporal cyclic effect, and semantic effect in a unified way using four bipartite graphs. POI-POI captures sequential effect, POI-Region graph captures geographical influence, POI-Time graph captures temporal cyclic effect and POI-Word graph captures semantic effect. The model embeds these four relational graphs into a shared low dimensional space. Then this model computes the similarity between a users' query (users' embedding, query time, and location) and the POIs that are not visited by that user. Most similar POIs are taken for the recommendation. 
        
        Liu et al. \cite{liu2017learning} proposed \textbf{SpatioTemporal Aware (STA)} which generalizes knowledge Graph Embedding \cite{lin2015learning} in their model. GE\cite{xie2016learning} embeds both users and POIs in a common latent space. The users and POIs are inherently different objects so that approach is unnatural. On the other hand, STA takes location and time as a spatiotemporal pair $<time, location>$ and uses the embedding of this pair as a relationship for connecting users and POIs.
        
        Christoforidis et al. \cite{christoforidis2018recommendation} proposed \textbf{Jointly Learn the Graph Embeddings (JLGE)} which uses six informational graphs. They are two unipartite (user-user and POI-POI) and four bipartite (user-location, user-time, location-user, and location-time). This model consists of a three-step process. In the first step, the model builds information graphs and weights the edges. In the second step, the model jointly learns the embeddings of the users and the POIs into the same latent space from these six informational graphs using the LINE model\cite{tang2015line}. Finally, in the third step, the model personalizes the POI recommendations for each user by tuning the influence of the participation networks for the final suggestions of the target user.

        The authors of the JLGE\cite{christoforidis2018recommendation} model extended their work by introducing a new model called \textbf{Recommendations with multiple Network Embeddings (RELINE)} \cite{christoforidis2019reline}. The model introduces two new networks: $i)$ Stay Points, which represents the locations of the user stayed the most, and $ii)$ Routes, the path followed when visiting POIs. Additionally two new bipartite graphs i.e., user-route and POI-stay points are added with the previous JLGE\cite{christoforidis2018recommendation} model for better capturing the users' preference dynamics.
        
        In another work, Xiong et al. \cite{xiong2020dynamic} proposed a semi-supervised learning framework called \textbf{Dynamic Spatio-temporal POI recommendation (DYSTAL)}. DYSTAL has two key components: a network embedding method and a dynamic factor graph model. Network embedding method jointly learns the embedding vectors of users and POIs of three subgraphs i.e., POI-POI, user-POI, and user-user to excavate complex spatio-temporal patterns of visiting behaviors. The Dynamic factor graph model captures different factors including the correlation of users’ vectors and POIs’ vectors from the previous embedding layer via the Factor Graph Model (FGM) \cite{lin2017detecting}. This model also considers the textual reviews of users by using SentiStrength \cite{xiong2018emotional} tool.
        
        Zhang et al.\cite{zhang2020modeling} propose \textbf{Hierarchical Category Transition (HCT)} which extends the Skip-Gram \cite{mikolov2013efficient} model to learn the hierarchical dependencies between POIs and categories, and the hierarchical category transition. HCT models the dynamic user preference by considering recently visited POIs and the associated hierarchical categorical sets. They formulated the dynamic user representation by incorporating the representations of POIs as well as the associated hierarchical category sets. Besides these models, Chen et al.\cite{chen2020modeling} uses embeddings and context filtering for modeling spatial trajectories. And Qiao et al.\cite{qiao2020heterogeneous} uses embeddings for predicting next POIs.
        
    \subsection{GAN}
    \label{GANmodel}
        Generative Adversarial Networks (GAN)\cite{goodfellow2014generative} is a popular model in deep learning where two Neural Network models compete with each other for giving better predictions. A very few papers use GAN in POI recommendation because POI recommendation problems do not necessarily fall under the solvable problem domain of GAN.
        
        Liu et al.\cite{liu2019geo} proposed \textbf{Geographical information-based adversarial learning model (Geo-ALM)} model that uses two modules: discriminator and generator which are essentially different but inspired from the conventional Generative Adversarial Network. The pairwise ranking is regarded as a discriminator that tries to predict the ranking relationship between generated sample pairs and is trained to maximize ranking samples’ likelihood. The generator continually generates critical negative samples, which are then coupled with positive samples, forming training instances. The framework interchangeably learns the parameters between two different modules.
        
        Zhou et al.\cite{zhou2019adversarial} proposed \textbf{Adversarial POI Recommendation (APOIR)} which combines GAN, GRU, and Matrix factorization for POI recommendation. GRU and MF combinedly learn both temporal and sequential preference of users. Two competitive components: recommender and discriminator is alternatively optimized by training both of them through an objective function using the learned preferences of users. The discriminator tries to maximize the probability of correctly distinguishing the true check-in locations from the generated recommended POIs by the recommender. Gao et al.\cite{gao2020adversarial} also uses GAN networks for identifying individuals by exploiting their trajectories.
    
    
    \subsection{Other models}
    \label{OTHERmodel}
        There are some of the other deep learning POI recommendation models that use hybrid architectures. In this section, we cover these methods. 
        
        Zhao et al.\cite{zhao2017geo} proposes \textbf{Geo-Temporal sequential embedding rank (Geo-Teaser)} which is based on the Skip-Gram \cite{mikolov2013efficient} model which learns the representations of context POIs given a target POI. The model attempts to learn the temporal POI embeddings through maximizing an objective function. The geographically hierarchical pairwise preference ranking model uses Bayesian Personalized ranking to learn the user preference on POIs. The core Geo-Teaser model is a unified framework that combines the temporal embedding model and the pairwise ranking method.
        
        Chang et al.\cite{chang2018content} proposed a somewhat different approach called \textbf{Content-Aware hierarchical POI Embedding (CAPE)} for POI recommendation. Most models do not use the text content of POIs, since most of the datasets do not contain such textual content. The authors thus generated a new dataset from Instagram which contains a textual description of POI written by the users. CAPE actually is a POI embedding model that consists of a check-in context layer and text content layer. The check-in content layer captures the geographical influence of POIs, while the text content layer captures the characteristics of POIs.
        
        A different type of work on missing POI check-in identification by Xi et al.\cite{xi2019modelling} is also notable in this context. Here, the authors proposed a model called \textbf{Bi-directional Spatio-Temporal Dependence and users’ Dynamic Preferences (Bi-STDDP)} to capture complex global spatial information, local temporal dependency relationships and users’ dynamic preferences. Bi-STDDP takes two check-in lists as input; one list for before the missing check-in, and one for after. Besides, this model also uses user-embedding, POI-embedding, and temporal patterns in the model.
        
        Zhou et al.\cite{zhou2019topic} proposed a hybrid architecture called \textbf{Topic-Enhanced Memory Network (TEMN)}. TEMN consists of three key parts: a Memory Network (MN), Temporal Latent Dirichlet Allocation (TLDA) \cite{zhou2018discovering} and geographical modeling. MN learns the complex interaction between user and POIs and captures neighbourhood-based interests of a user. TLDA is an unsupervised generative probabilistic model which captures temporal preferences and inner interest of users. This module provides the pattern-user probability distribution.  The distributions of venues and time slots associated with each pattern can also be estimated through TLDA. TLDA and MN can jointly learn characteristics of users and POIs. The Geographical model captures the geographical influence.
        
        Recent researches show that hierarchical structures can be modeled using hyperbolic representation methods \cite{de2018representation, nickel2017poincare, nickel2018learning}. For that reason, Feng et al. \cite{feng2020hme} propose a novel \textbf{Hyperbolic Metric Embedding (HME)} approach for the next-POI recommendation task. HME can be divided into two part i.e., (1) Hyperbolic Metric Embedding: uses Poincaré ball model \cite{nickel2017poincare} to learn four different relationships (POI-POI, POI-User, POI-Category, and POI-Region) by projecting them in a shared hyperbolic space; (2) Recommending with Hyperbolic Embeddings: combines the user preferences and POI sequential transitions in the Poincaré ball model an Einstein midpoint aggregation method \cite{gulcehre2018hyperbolic, ungar2005analytic}. The geographical distance is also considered in this model because users tend to visit the POI that are close to them \cite{feng2015personalized}.
        
        Along with the aforementioned models, recent works have proposed some advanced models that take advantage of multiple techniques to improve performance. Zhang et al.\cite{zhang2020personalized} use embeddings for geographical influence modeling. Wang et al.\cite{wang2017lce} use embeddings for predicting the next POIs. Ding et al.\cite{ding2018spatial} use DNN for time-specific POI recommendation. The CNN has been studied in terms of POI recommendation in some prior works \cite{xu2019ssser, chen2020cem}. On the other hand, Massimo et al.\cite{massimo2018harnessing} also experimented with Inverse Reinforcement Learning (IRL) \cite{ramachandran2007bayesian, neu2009training, ziebart2008maximum} to analyze the performance of IRL in POI recommendation.\\
        
        
   \noindent 
   \textbf{Summary of different paradigms of deep learning models:}\\
   By analyzing all the aforementioned models, we can get a comprehensive picture of how different deep learning paradigms have been utilized to handle different aspects of POI recommendation models. Since successful recommendation fundamentally depends on historical POI information, sequential models (i.e., RNN, LSTM) have been primarily used in recent POI recommendation models. Among them, LSTM is the most popular approach due to its long term sequential information capturing capability. A large body of works has experimented with minor modifications of LSTMs to improve long and short-term modeling preferences. Thus, sequence information capture has been the center of interest in POI recommendation models. Recent state-of-the-art models are leveraging the self-attention transformer mechanism for POI recommendation, which greatly suppresses the problems associated with very long sequences. They are also computationally parallelizable. Besides capturing historical sequence information, researchers have also analyzed the spatial influence in this regard. Due to the nature of check-in data, users and POIs form a relationship graph that can be highly useful to model the spatial dependency. Thus, graph embedding methods have garnered attention in the past couple of years in POI recommendation. LSTM coupled with these graph embedding models thus opens up the opportunity to capture both sequential and spatiotemporal features from given data. Advanced models have also utilized adversarial learning models and specialized embedding methods to achieve state-of-the-art performance results.
        
\newcommand\dnnCount{3}
\newcommand\rnnCount{2}
\newcommand\basiclstmCount{3}
\newcommand\bilstmCount{1}
\newcommand\modifiedlstmCount{1}
\newcommand\lstmCount{5}
\newcommand\ganCount{1}
    
\begin{table*}[!t]
\small
\begin{threeparttable}
\caption{Categorization of Papers}
\label{paperModel}
\centering
\begin{tabular}{ |c|c|l|p{120mm}|}
    \hline
    \multicolumn{1}{|c}{\textbf{Category}} & 
    \multicolumn{1}{|c}{\textbf{Subcategory}} & 
    \multicolumn{1}{|c}{\textbf{Year}} & 
    \multicolumn{1}{|c|}{\textbf{Reference}} \\
    
    \hline
    \multirow{3}{*}{Basic RNN} & \multirow{3}{*}{-} & 2016 & ST-RNN\cite{liu2016predicting}\\
    \cline{3-4}
    & & 2018 & MCI-DNN\cite{liao2018multi}\\
    \cline{3-4}
    & & 2020 & Flashback\cite{yang2020location}, ASPPA\cite{zhao2020discovering}, LLRec\cite{wang2020next}, DRLM\cite{huang2020deep}\\
    \hline
    
    \multirow{6}{*}{LSTM} & \multirow{3}{*}{Basic LSTM} & 2018 & HST-LSTM\cite{kong2018hst}, TMCA\cite{li2018next}, SLSTM\cite{zhao2018personalized}\\
    \cline{3-4}
     & & 2019 & LSPL\cite{wu2019long}, ASTEN\cite{doan2019attentive}, ATST-LSTM\cite{huang2019attention}, MMR\cite{li2019multi}, SGBA\cite{yang2019next}\\
     \cline{3-4}
     & & 2020 & PLSPL\cite{wu2020personalized}, LSTPM\cite{sun2020go}, iMTL\cite{zhang2020interactive}, CatDM\cite{yu2020category}, ARNN\cite{guo2020attentional}, STAR\cite{wang2020modeling}\\
    \cline{2-4}
    
     & \multirow{1}{*}{Bi-LSTM} & 2020 & GT-HAN\cite{liu2020exploiting}, t-LocPred\cite{liu2020time}, CAPRE\cite{chang2020content}\\
    \cline{2-4}
    
     & \multirow{1}{*}{Modified LSTM} & 2020 & STGN\cite{zhao2019stgn}\\
    \cline{2-4}
    
    & \multirow{1}{*}{Self-Attention} & 2020 & GeoSAN\cite{lian2020geography}, SANST\cite{guo2020sanst}\\
    \hline
    
    \multirow{2}{*}{GRU} & \multirow{2}{*}{-} & 2018 & DeepMove\cite{feng2018deepmove}, CARA\cite{manotumruksa2018contextual} \\
    \cline{3-4}
    & & 2019 & MGRU\cite{kala2019context} \\
    \hline
    
    \multirow{4}{*}{Graph Embedding} & \multirow{4}{*}{-} & 2016 & GE\cite{xie2016learning}\\
    \cline{3-4}
    & & 2017 & STA\cite{liu2017learning} \\
    \cline{3-4}
    & & 2018 & JLGE\cite{christoforidis2018recommendation}\\
    \cline{3-4}
    & & 2019 & RELINE\cite{christoforidis2019reline} \\
    \cline{3-4}
    & & 2020 & DYSTAL\cite{xiong2020dynamic}, HCT\cite{zhang2020modeling}, UP2VEC\cite{qiao2020heterogeneous}, HMRM\cite{chen2020modeling} \\
    \hline
    
    \multirow{2}{*}{GAN} & \multirow{2}{*}{-} & 2019 & Geo-ALM\cite{liu2019geo}, APOIR\cite{zhou2019adversarial}\\
    \cline{3-4}
    & & 2020 & AdattTUL\cite{gao2020adversarial}\\
    \hline
    
    \multirow{4}{*}{Others} & \multirow{4}{*}{-} & 2017 & Geo-Teaser\cite{zhao2017geo}, LCE\cite{wang2017lce}\\
    \cline{3-4}
    & & 2018 & CAPE\cite{chang2018content}, ST-DME\cite{ding2018spatial}\\
    \cline{3-4}
    & & 2019 & Bi-STDDP\cite{xi2019modelling}, TEMN\cite{zhou2019topic}, SSSER \cite{xu2019ssser}\\
    \cline{3-4}
    & & 2020 & HME\cite{feng2020hme}, MPR\cite{luo2020spatial}, PGIM\cite{zhang2020personalized}, CEM \cite{chen2020cem}\\
    \hline
\end{tabular}
\end{threeparttable}
\end{table*}

\begin{table*}[!t]
\small
\begin{threeparttable}
\caption{Model summery and Performance of the state-of-the-art method}
\label{paperComapreTable}
\centering
\begin{tabular}{ p{21mm}|p{87mm}|p{72mm}}
    \hline
    \textbf{Model, Venue year [ref]} & \multicolumn{1}{|c}{\textbf{Method Summery}} & \multicolumn{1}{|c}{\textbf{Performance}}\\
    \hline
    ST-RNN, \textit{AAAI 2016} \cite{liu2016predicting} & Use time-specific and distance-specific transition matrix for capturing temporal cyclic effect and geographical influence respectively. & Gowalla: Rec@5\tnote{1}=0.1524, Rec@10=0.2714.\newline GTD: Rec@5=0.4986, Rec@10=0.6812\\
    \hline
    Flashback, \textit{IJCAI 2020} \cite{yang2020location} & Model sparse user mobility data by doing flashbacks on hidden states in RNNs and uses the weighted average of historical hidden states for better capturing spatio-temporal effects. & Foursquare: Acc\tnote{2}@5=0.5399, Acc@10=0.6236.\newline Gowalla: Acc@5=0.2754, Acc@10=0.3479.\\
    \hline 
    ASPPA, \textit{IJCAI 2020} \cite{zhao2020discovering} & Automatically identify the semantic subsequnce of POIs and discovers their sequential patterns by hierarchically learning the latent structure from check-in list and power-law attention mechanism. & Foursquare (US): Acc@10=0.3371, Acc@20=0.3950.\newline Gowalla: Acc@10=0.2947, Acc@20=0.3573.\\
    \hline
    LLRec (Teacher), \textit{WWW 2020} \cite{wang2020next} & Capture long-term, short-term preferences, textual feature of POIs and complex dependencies among user preferences  by using embedding, recurrent component and attention mechanism. & Foursquare: Acc@10=0.3542, Acc@20=0.4594 \newline Gowalla: Acc@10=0.3874, Acc@20=0.4781\\
    \hline
    
    HST-LSTM, \textit{IJCAI 2018} \cite{kong2018hst} & Use hierarchical model using LSTM to encode the periodicity of people's movement. & Baidu Map:  Acc@10=0.4847, Acc@20=0.5657\\
    \hline
    TMCA, \textit{ICDM 2018} \cite{li2018next} & Capture complex spatial and temporal dependencies among historical check-in activities by using LSTM based encoder-decoder model, attention mechanism and embedding method. & Gowalla: Rec@5=0.21926, Rec@10=0.27725.\newline Foursquare: Rec@5=0.02870, Rec@10=0.04809.\\
    \hline 
    LSPL, \textit{CIKM 2019} \cite{wu2019long} & Capture both sequential and contextual information via long-term and short-term preference learning. & Foursquare (NYC): Prec\tnote{3}@10=0.3901, Prec@20=0.4461 \newline Foursquare (TKY): Prec@10=0.3986, Prec@20=0.4596.\\
    \hline
    PLSPL, \textit{TKDE 2020} \cite{wu2020personalized} & Extend their previous work LSPL \cite{wu2019long} by introducing user-based linear combination unit which better captures user preferences. & Foursquare (NYC): Prec@10=0.3953, Prec@20=0.4475 \newline Foursquare (TKY): Prec@10=0.4020, Prec@20=0.4664.\\
    \hline
    ASTEN, \textit{PAKDD 2019} \cite{doan2019attentive} & Capture the sequential, temporal and geographical influence by using LSTM and attention mechanism. & Foursquare (US): Rec@5=0.328, Rec@10=0.414 \newline Foursquare (EU): Rec@5=0.281, Rec@10=0.35 \newline Gowalla: Rec@5=0.152, Rec@10=0.266 \\
    \hline
    LSTPM, \textit{AAAI 2020} \cite{sun2020go} & Capture long-term preference modeling by using a non-local network and short-term preference modeling by using geo-dialated LSTM. & Foursquare (NY): Rec@5=0.3372, Rec@10=0.4091 \newline Gowalla: Rec@5=0.2021, Rec@10=0.2510\\
    \hline
    iMTL, \textit{IJCAI	2020} \cite{zhang2020interactive} & Use a two-channel encoder and a task-specific decoder for capturing the sequential correlations of activities and location preferences. & POI Prediction: \newline Foursquare (CLT): Rec@10=0.0534, Map\tnote{4}@10=0.0238 \newline  Foursquare (CAL): Rec@10=0.0691, Map@10=0.0443 \newline Foursquare (PHO): Rec@10=0.0769, Map@10=0.0352 \\
    \hline
    CatDM, \textit{WWW 2020} \cite{yu2020category} & capture temporal influence, geographical influence and overcome data sparsity by using two LSTM based deep encoder, two filter, metric embedding and attention mechanism. & Foursquare (NYC): Rec@5=0.2407, Rec@10=0.3113 \newline Foursquare (TKY): Rec@5=0.2148, Rec@10=0.2739.\\
    \hline
    ARNN, \textit{AAAI 2020} \cite{wang2020next} & Capture data sparsity by using new concept called transition regularity. Also capture sequential, spatial, temporal, semantic influence by using embedding, knowledge graph, LSTM and attention mechanism. & Foursquare (NY): Acc@10=0.4162, Acc@20=0.4393 \newline Foursquare (TK): Acc@10=0.4285, Acc@20=0.4864 \newline Gowalla (SF): Acc@10=0.2336, Acc@20=0.2530\\
    \hline
    
    GT-HAN, \textit{Neurocomputing 2020} \cite{liu2020exploiting} & Capture great variation in geographical co-influence across POIs, temporal dependency and sequence dependency in check-in list by using embedding layer, Bi-LSTM and attention mechanism. & Foursquare: AUC\tnote{8}=0.9661, acc@5: 0.13-0.15, \newline acc@10: 0.17-0.19, acc@20: 0.23-0.25 \newline (depending on latent dimensionality)\\
    \hline 
    t-LocPred, \textit{TKDE 2020} \cite{liu2020time} & capture a users' coarse-grained spatiotemporal movement pattern by using CNN and ConvLSTM and fine-grained POI check-in information by using spatial-aware memory-augmented LSTM with time-aware attention. & Gowalla: MRR\tnote{5}=0.247 (C=6, all), \newline Weeplaces: MRR=0.277 (C=6, all), \newline Brightkite: MRR=0.388 (C=4, all)\\
    \hline
    CAPRE, \textit{SDM 2020} \cite{chang2020content} & Capture the various perspectives of user about POIs along with content-aware and geographical user behavior pattern by using character-level CNN, multi-head attention, Bi-LSTM and MLP. & Foursquare: Rec@5=0.1724, Rec@10=0.2084 \newline Instagram: Rec@5=0.2934, Rec@10=0.3588\\
    \hline
    STGCN, \textit{AAAI 2019} \cite{zhao2019stgn} & Modify the basic LSTM model slightly by introducing new gates and cell to capture short-term and long-term preference easily.& Foursquare (CA): Acc@5=0.1308, Acc@10=0.1612.\newline Foursquare (SIN): Acc@5=0.2737, Acc@10=0.3017.\newline Gowalla: Acc@5=0.1644, Acc@10=0.2020.\newline Brightkite: Acc@5=0.4953, Acc@10=0.5231.\\
    \hline
\end{tabular}
\begin{tablenotes}
    \item[1] Recall@K is the presence of the correct POI among the top K recommended POIs \cite{sun2020go}.\\
    \item[2] Acc@k is 1 if the visited POI appears in the set of top-K recommendation POIs and 0 otherwise \cite{zhao2019stgn}. The overall Acc@K is calculated as the average value of all testing instances. Also known as Accouracy@K or Hit Rate@k or Hit Ratio@k or HR@k.\\
    \item[3] Precision@K indicates that whether the ground truth POI appears in the top-k recommended POIs \cite{wu2019long}.\\
    \item[4] MAP (Mean Average Precision) measures the order of our recommendation list \cite{wu2019long}.\\
    \item[5] MRR is the average reciprocal rank of positive examples. This metric reflects the overall ranking ability of the model \cite{zhao2020discovering}.\\
\end{tablenotes}
\end{threeparttable}
\end{table*}

\begin{table*}[!t]
\small
\begin{threeparttable}
\centering
\begin{tabular}{ p{22mm}|p{86mm}|p{72mm}}
    \hline
    \textbf{Model, Venue year [ref]} & \multicolumn{1}{|c}{\textbf{Method Summery}} & \multicolumn{1}{|c}{\textbf{Performance}}\\
    \hline
    GeoSAN, \textit{KDD 2020} \cite{lian2020geography} & Resolve the sparsity issue by introducing a new loss function and represent the hierarchical gridding of each GPS point with a self-attention based geography encoder for better use of geographical information. & Foursquare: Acc@5=0.3735, Acc@10=0.4867.\newline  Gowalla: Acc@5=0.4951, Acc@10=0.6028.\newline Brightkite: Acc@5=0.5258, Acc@10=0.6425. \\
    \hline
    SANST, \textit{arXiv 2020} \cite{guo2020sanst} & Capture the spatial-temporal and sequential patterns by using embedding, self-attention network (transformer network), Bi-LSTM. & Gowalla: Acc@10=0.2273,\newline Los Angeles: Acc@10=0.3941,\newline Singapore: Acc@10=0.2417\\
    \hline
    
    DeepMove, \textit{WWW 2018} \cite{feng2018deepmove} & Capture the complex dependencies and multi-level periodicity nature of human by using embedding, GRU and attention mechanism. & Foursquare (NY): Rec@5=0.3372, Rec@10=0.4091. \newline Gowalla: Rec@5=0.2021, Rec@10=0.2510\\
    \hline
    CARA, \textit{SIGIR 2018} \cite{manotumruksa2018contextual} & Capture the different types of impact of different contextual information by using embedding, GRU and two gating mechanism. & Foursquare: Acc@10=0.8851, Yelp: Acc@10=0.5587,\newline Brightkite: Acc@10=0.7385\\
    \hline
    MGRU, \textit{JAIHC 2019} \cite{kala2019context} & Capture dynamic and transition context using Multi-GRU (Two special gate are added with GRU). & Foursquare: Rec@10=0.9214, Rec@15=0.9214 \newline Gowalla: Rec@10=0.8512, Rec@15=0.8765\\
    \hline
    
    GE, \textit{CIKM 2016} \cite{xie2016learning} & Capture data sparsity, context awareness, cold start, dynamic of personal preference by using the embedding of four graphs into a shared low dimensional space. & Foursquare: Acc@10=0.372, Acc@20=0.435. \newline Gowalla: Acc@10=0.462, Acc@20=0.533.\\
    \hline
    STA, \textit{arXiv 2017} \cite{liu2017learning} & This paper generalizes the knowledge Graph Embedding and takes location and time as a spatiotemporal pair for connecting users and POIs. & Foursquare: Acc@10=0.439, Acc@20=0.486. \newline Gowalla: Acc@10=0.488, Acc@20=0.540.\\
    \hline
    JLGE, \textit{DSAA 2018} \cite{christoforidis2018recommendation} & Jointly learn the embeddings of the users and the POIs into the same latent space from the six informational graphs using LINE model\cite{tang2015line}. & Foursquare: Acc@10=0.410, Acc@20=0.462. \newline Weeplaces: Acc@10=0.488, Acc@20=0.536.\\
    \hline
    RELINE, \textit{arXiv 2019} \cite{christoforidis2019reline} & This paper extends the previously discussed paper \cite{christoforidis2018recommendation} by adding two new networks: i.e., stay points and routes. & Foursquare: Acc@10=0.410, Acc@20=0.462.\newline Weeplaces: Acc@10=0.488, Acc@20=0.536.\newline Gowalla: Acc@10=0.518, Acc@20=0.556.\\
    \hline
    Geo-Teaser, \textit{WWW 2017} \cite{zhao2017geo} & Use Skip-Gram model for temporal POI embedding and Bayesian Personalized Ranking for pairwise ranking of POIs. A unified framework combines the temporal POI embedding and pairwise ranking model. & Foursquare: Prec@5=0.13, Prec@10=0.1, \newline Rec@5=0.15, Rec@10=0.2\newline Gowalla: Prec@5=0.16, Prec@10=0.13, \newline Rec@5=0.07, Rec@10=0.12\\ 
    \hline
    DYSTAL, \textit{Information Processing and Management 2020} \cite{xiong2020dynamic} & Capture complex spatio-temporal patterns of visiting behaviors by jointly learning the effects of users’ social relationships, textual reviews, and POIs’ geographical proximity using a network embedding method and dynamic factor graph model. & Foursquare (SIN): Acc@10=0.232, Rec@10=0.152, \newline Yelp: Acc@10=0.206, Rec@10=0.098\\
    \hline
    HCT, \textit{Information Sciences, Elsevier 2020}\cite{zhang2020modeling} & Utilize Skip-Gram model to model the categorical transitions at different layers of categorical hierarchies as well as the hierarchical dependencies between POIs and categories & Foursquare(SIN): Prec@5=0.613 Rec@5=0.0403\newline Foursquare(NYC): Prec@5=0.0585, Rec@5=0.0352\newline Foursquare(LA): Prec@5=0.0653, Rec@5=0.0305 \\
    \hline
    
    APOIR, \textit{2019} \cite{zhou2019adversarial} & Use matrix factorization and GRU to learn user preferences and train two competitive component: recommender and discriminator to generate prediction. & Yelp: Prec@5=0.1, Rec@5=0.16, MAP@5=0.233, NDCG\tnote{6}@5=0.094\\
    \hline
    
    CAPE, \textit{IJCAI 2018} \cite{chang2018content} & Use text content layer and check-in content layer for embeds the POIs and generates their own dataset. & With STELLAR: Rec@5=0.2384, Rec@10=0.2989 \newline With LSTM: Rec@5=0.2412, Rec@10=0.3054 \newline With GRU: Rec@5=0.2433, Rec@10=0.3079 \newline With ST-RNN: Rec@5=0.2239, Rec@10=0.2601 \\
    \hline
    TEMN, \textit{KDD 2019} \cite{zhou2019topic} & Capture both neighbourhood-based and global preferences by using a combination of supervised and unsupervised learning. & WeChat (GPR): \newline
    TEMN (GPR): Acc@5=0.70389, Acc@10=0.81752.\newline TEMN (CPR): Acc@5=0.72876, Acc@10=0.83398.\\
    \hline
    HME, \textit{SIGIR 2020} \cite{feng2020hme} & Capture  POI sequential transitions, geographical, semantic and user preferences by using hyperbolic metric embedding along with Poincaré ball and Einstein midpoint aggregation method. & Foursquare (NYC): Rec@5=0.0962, Rec@10=0.1371 \newline Foursquare (TKY): Rec@5=0.1527, Rec@10=0.2172. \newline Gowalla (Houston): Rec@5=0.1533, Rec@10=0.2318\\
    \hline
\end{tabular}
\begin{tablenotes}
    \item[6] NDCG (Normalized Discounted Cumulative Gain) measures the quality of top-K ranking list \cite{sun2020go}.\\
\end{tablenotes}
\end{threeparttable}
\end{table*}

\begin{table*}[!t]
\small
\begin{threeparttable}
\caption{Descriptions of Datasets}
\label{datasetModel}
\centering
\begin{tabular}{ |p{15mm}|p{40mm}|p{30mm}|p{20mm}|p{15mm}|p{15mm}|p{15mm}|}
    \hline
    \textbf{Dataset name} & 
    \textbf{Reference} &
    \textbf{Date} & 
    \textbf{Region} &
    \textbf{\#User} & 
    \textbf{\#POI} & 
    \textbf{\#Check-in}\\
    %
    \hline
    \multirow{26}{*}{Foursquare} 
    & Flashback\cite{yang2020location} & Apr 2012 - Jan 2014 & World & 46065 & 69005 & 9450342 \\
    \cline{2-7}
    & ASPPA\cite{zhao2020discovering} & Apr 2012 - Sep 2013 & US (except Alaska, Hawaii) & 49005 & 206097 & 425691\\
    \cline{2-7}
    & LSPL\cite{wu2019long}, PLSPL\cite{wu2020personalized} & \multirow{2}{*}{Apr 2012 - Feb 2013} 
    & New York & 1083 & 38333 & 227428 \\
    \cline{4-7}
    & CatDM\cite{yu2020category}, ARNN\cite{guo2020attentional}, Bi-STDDP\cite{xi2019modelling}, HME\cite{feng2020hme} 
    & & Tokyo & 2293 & 61858 & 573703 \\
    \cline{2-7}
    & \multirow{2}{*}{ASTEN\cite{doan2019attentive}} & \multirow{2}{*}{-} & USA & 21878 & 21651 & 569091\\
    \cline{4-7}
    & & & EU & 15387 & 115567 & 3227845\\
    \cline{2-7}
    & LSTPM\cite{sun2020go} & Feb 2010 - Jan 2011 & New York & 934 & 9296 & 52983 \\
    \cline{2-7}
    & \multirow{3}{*}{iMTL\cite{zhang2020interactive}} 
    & \multirow{3}{*}{Apr 2012 - Sep 2013} & Charlotte & 1580 & 1791 & 20940 \\
    \cline{4-7}
    & & & Calgary & 301 & 985 & 13954 \\
    \cline{4-7}
    & & & Phoenix & 1623 & 2441 & 22620 \\
    \cline{2-7}
    & GT-HAN\cite{liu2020exploiting}, APOIR\cite{zhou2019adversarial} & Apr 2012 - Sep 2013 & USA & 24941 & 28593 & 1196248 \\ 
    \cline{2-7}
    & CAPRE\cite{chang2020content} & - & - & 4163 & 121142 & 483813\\
    \cline{2-7}
    & \multirow{2}{*}{STGN\cite{zhao2019stgn}} & Jan 2010 - Feb 2011 & California & 49005 & 206097 & 425691\\
    \cline{3-7}
    & & Aug 2010 - Jul 2011 & Singapore & 30887 & 18995 & 860888 \\
    \cline{2-7}
    & GeoSAN\cite{lian2020geography} & Apr 2012 - Jan 2014 & World & 12695 & 37344 & 1941959 \\
    \cline{2-7}
    & DeepMove\cite{feng2018deepmove} & Feb 2010 - Jan 2011 & New York & 15639 & 43379 & 293559\\
    \cline{2-7}
    & CARA\cite{manotumruksa2018contextual} & - & World & 10766 & 10695 & 1336278 \\
    \cline{2-7}
    & MGRU\cite{kala2019context} & Aug 2010 - Jul 2011 & Singapore & 4630 & 6176 & 201525\\
    \cline{2-7}
    & GE\cite{xie2016learning} & Sep 2010 - Jan 2011 & World & 114508 & 114,508 & 1434668 \\
    \cline{2-7}
    & STA\cite{liu2017learning}, JLGE\cite{christoforidis2018recommendation}, RELINE\cite{christoforidis2019reline} & Sep 2010 - Jan 2011 & USA & 114508 & 62462 & 1434668\\ 
    \cline{2-7}
    & Geo-ALM\cite{liu2019geo} & Aug 2010 - Jul 2011 & Singapore & 2321 & 5596 & 194108 \\ 
    \cline{2-7}
    & DYSTAL\cite{xiong2020dynamic} & - & Singapore & 74250 & - & -\\
    \cline{2-7}
    & \multirow{3}{*}{HCT\cite{zhang2020modeling}} & \multirow{3}{*}{-} & Singapore & 2676 & 1633 & 116757 \\
    \cline{4-7}
    & & & New York City & 1982 & 2454 & 187750\\
    \cline{4-7}
    & & & Los Angeles & 2109 & 1576 & 70189\\
    \cline{2-7}
    & Geo-Teaser\cite{zhao2017geo} & Jan 2011 - Jul 2011 & World & 10034 & 16561 & 865647\\
    %
    \hline
    \multirow{16}{*}{Gowalla}
    & ST-RNN\cite{liu2016predicting} & Feb 2009 - Oct 2010 & World & 10997 & - & 6400000 \\ 
    \cline{2-7}
    & Flashback\cite{yang2020location} & Feb 2009 - Oct 2010 & World & 52979 & 121851 & 3300986 \\
    \cline{2-7}
    & ASPPA\cite{zhao2020discovering} & Feb 2009 - Oct 2010 & World & 4996 & 6871 & 245157  \\
    \cline{2-7}
    & LLRec\cite{wang2020next}, RELINE\cite{christoforidis2019reline} & Jan 2009 - Aug 2011 & World & 319063 & 2844076 & 36001959 \\
    \cline{2-7}
    & TMCA\cite{li2018next} & Feb 2009 - Oct 2010 & World & 22209 & 50569 & 1493799 \\
    \cline{2-7}
    & ASTEN\cite{doan2019attentive} & Feb 2009 - Oct 2010 & World & 52484 & 115567 & 3227845\\
    \cline{2-7}
    & LSTPM\cite{sun2020go} & Feb 2009 - Oct 2010 & World & 5802 & 40868 & 301080 \\
    \cline{2-7}
    & ARNN\cite{guo2020attentional} & Feb 2009 - Oct 2010 & San Fransisco & 170 & 7340 & 32058 \\
    \cline{2-7}
    & GT-HAN\cite{liu2020exploiting}, STGN\cite{zhao2019stgn}, APOIR\cite{zhou2019adversarial} & Feb 2009 - Oct 2010 & World & 18737 & 32510 & 1278274 \\
    \cline{2-7}
    & t-LocPred & Jan 2009 - Oct 2010 & Goteborg (Sweden) & 5342 & 12229 & 103787\\
    \cline{2-7}
    & GeoSAN\cite{lian2020geography} & - & - & 31708 & 131329 & 2963373 \\
    \cline{2-7}
    & SANST\cite{guo2020sanst}, Geo-ALM\cite{liu2019geo} & Feb 2009 - Oct 2010 & California, Nevada & 10162 & 24250 & 456988 \\
    \cline{2-7}
    & MGRU\cite{kala2019context} & Nov 2009 - Oct 2010 & Austin & 2321 & 5596 & 194108\\
    \cline{2-7}
    & GE\cite{xie2016learning}, STA\cite{liu2017learning}, Bi-STDDP\cite{xi2019modelling} & Feb 2009 - Oct 2010 & World & 107092 & 1280969 & 6442892 \\
    \cline{2-7}
    & Geo-Teaser\cite{zhao2017geo} & Jan 2011 - Jul 2011 & World & 3240 & 33578 & 556453 \\
    \cline{2-7}
    & HME\cite{feng2020hme} & Nov 2010 - Jun 2011 & Houston & 4627 & 15135 & 362783 \\
    \hline
    %
    \multirow{4}{*}{Brightkite}
    & t-LocPred\cite{liu2020time} & May 2008 - Oct 2010 & Tokyo & 2263 & 37196 & 183298\\
    \cline{2-7}
    & STGN\cite{zhao2019stgn} & May 2008 - Oct 2010 & World & 51406 & 772967 & 4747288 \\
    \cline{2-7}
    & GeoSAN\cite{lian2020geography} & Apr 2008 - Oct 2010 & World & 5247 & 48181 & 1699579 \\
    \cline{2-7}
    & CARA\cite{manotumruksa2018contextual} & Apr 2008 - Oct 2010 & World & 14374 & 5050 & 681024 \\
    \hline
    %
    \multirow{2}{*}{WeChat}
    &  \multirow{2}{*}{TEMN\cite{zhou2019topic}} & \multirow{2}{*}{Sep 2016 - Aug 2017} & \multirow{2}{*}{Beijing} & 28566 & 13826 & 509589 \\
    \cline{5-7}
    & & & & 75973 & 28183 & 5644965 \\
    \hline
\end{tabular}
\end{threeparttable}
\end{table*}

\begin{table*}[!t]
\small
\begin{threeparttable}

\centering
\begin{tabular}{ |p{15mm}|p{40mm}|p{30mm}|p{20mm}|p{15mm}|p{15mm}|p{15mm}|}
    \hline
    \textbf{Dataset name} & 
    \textbf{Reference} &
    \textbf{Date} & 
    \textbf{Region} &
    \textbf{\#User} & 
    \textbf{\#POI} & 
    \textbf{\#Check-in}\\
    
    \hline
    \multirow{4}{*}{Yelp}
    & TMCA\cite{li2018next} & Jan 2014 - Jun 2017 & World & 11564 & 18683 & 492489 \\
    \cline{2-7}
    & CARA\cite{manotumruksa2018contextual} & - & World & 38945 & 34245 & 981379 \\
    \cline{2-7}
    & DYSTAL\cite{xiong2020dynamic} & - & Las Vegas & 337084 & 26809 & 1605396\\
    \cline{2-7}
    & APOIR\cite{zhou2019adversarial} & - & World & 30887 & 18995 & 860888 \\
    \hline
    \multirow{1}{*}{Baidu}
    & HST-LSTM\cite{kong2018hst} & Dec 2015 - Dec 2015 & Peking & - & - & -\\
    
    \hline
    \multirow{1}{*}{Instagram}
    &  CAPRE\cite{chang2020content}, CAPE\cite{chang2018content} & - & New York City & 78233 & 13187 & 2216631 \\
    
    \hline
    \multirow{2}{*}{Weeplaces}
    & LLRec\cite{wang2020next}, JLGE\cite{christoforidis2018recommendation}, RELINE\cite{christoforidis2019reline} & Nov 2003 - Jun 2011 & World & 15799 & 971309 & 7658368 \\
    \cline{2-7}
    & t-LocPred\cite{liu2020time} & Nov 2003 - Jun 2011 & New York City & 4855 & 38537 & 900906\\
    \hline
\end{tabular}
\end{threeparttable}
\end{table*}

\section{POI Sequence Recommendation}

\label{section:poisequence}
        So far we've primarily discussed the next POI recommendation models. However, some state-of-the-art methods also recommend a sequence of POIs that are more likely to be visited by the user in the future. In POI sequence recommendation the input is a sequence of check-ins or check-in list and the output also a check-in list. So, we can think of this recommendation as a sequence to sequence (Seq2Seq) task. Several state-of-the-art methods have been proposed in recent years for solving the POI sequence recommendation task. Here we briefly discuss some of them.\\
        
        Baral et al. \cite{baral2018close} proposed \textbf{Contextualized Location Sequence Recommender (CLoSe)} which incorporates different contexts (e.g., social, temporal, categorical, and spatial) into the hidden and output layer. This model uses either the simple RNN or the LSTM model. The results show that CLoSe-LSTM performs better than CLoSe-RNN.
        
        Huang et al. \cite{huang2019dynamic} proposed \textbf{Dynamic Recommendation of POI Sequence (DRPS)} which is based on DNN. This model consists of an encoder and a decoder module and for getting better performance, this model takes into account the POI embedding feature, the geographical and categorical influences of historical trajectories, and the positional encoding. This paper also proposed two new evaluation metrics for better performance evaluation.
         
        In another work, Lu et al. \cite{lu2020glr} proposed \textbf{Graph-based Latent Representation model (GLR)} which can capture geographical influence, temporal influence, user preference, etc. GLR learns the latent vectors based on word2vec \cite{le2014distributed} technique. Here, the authors added user preference, temporal successive transition influence, geographic influence, and LSTM \cite{lstmarticle} with GLR model and propose a new model GLR\_GT\_LSTM which can capture users’ complex successive transition behavior.
        
        Alongside these models, Wang et al. \cite{wang2019spent}, Baral et al. (HiCaPS) \cite{baral2018hicaps}, Lin et al. \cite{lin2018successive} also proposed different POI sequence recommendation models. On the other hand, Li et al. \cite{li2019context} worked with a variation of this task, missing POI check-ins prediction by leveraging an attention-based seq2seq generative model.

\section{Influential Factors}

\label{section:influentialfactors}

In Section \ref{section:recommendationmodels}, we have highlighted that extensive research has been done in the domain of POI recommendation. In all these works, researchers have tried to figure out the most influential factors that affect Point-of-interest recommendation. It is quite difficult to conclude the factors of choosing a next POI as POI recommendation is greatly affected by human behavior which changes over time. However, most researchers agree with some common factors like sequential effect, geographical influence, semantic effect, social influence, temporal influence, etc. that affect POI recommendations. These factors are derived from the behavior of the users' decisions and we need to capture these influential factors in our models for a better recommendation of POI. In this section, we elaborately discuss these influential factors. In Table \ref{influenceModel}, we also summarize how different POI recommendation models cover these influential factors.
    \subsection{Sequential Effect}
        The sequential effect of POI recommendation is similar to the analogy of NLP problems- such as making a sentence where the next word depends on the previous words. \\
        \cite{cheng2013you, zhang2014lore, xie2016learning, christoforidis2019reline} also show that sequential effect puts a major impact on POI recommendation. Zhang et al.\cite{zhang2014lore} extracts sequential patterns from the location sequences of all users and model them as a concise location-location transition graph. Then they determine the transition probabilities in terms of transition counts and outgoing counts. Finally, the model processes the check-in locations according to their arrival order and incrementally updates the constructed location-location transition graph. The sequential probabilities are derived with additive Markov Chain applied on the location-location graph. Xie et al. \cite{xie2016learning} designed a fully connected deep LSTM network for skeleton-based action recognition. This architecture enables fully exploit the inherent correlations among skeleton joints to capture sequential effect. Christoforidis et al. \cite{christoforidis2019reline} jointly learns the graph embeddings of different information networks in the same latent space. The model is optimized using negative sampling. All the embeddings of the input bipartite graphs are integrated into the model.
    \subsection{Geographical Influence}
        Recent researches show that people tend to visit places that are close to him/her or are close to the places already visited by that person. Users who check in a location within a region have a relatively larger probability to visit the places in close proximity. Users tend to go to stores, marketplaces, or visit picnic spots that are close to where they live. Also, after visiting a certain tourist spot, people tend to go to nearby restaurants or malls. Thus, spatial proximity is a worthy concern to predict users' next location.\\
        \cite{ye2011exploiting, xie2016learning, sun2020go, kefalas2017time, liu2013learning, chang2018content} show great interest in analyzing the geographical influence for predicting POIs. Ye et al. \cite{ye2011exploiting} perform spatial analysis on real datasets of Foursquare and Whrrl. The study finds the implication of distance on user check-in behavior by measuring the probability of a pair of check-ins being within a certain distance. The study confirmed the above-mentioned implications of the proximity of POI in predicting the next POI. The model introduces a collaborative recommendation method based on the naive Bayesian method to realize the POI recommendation. The paper proposes a unified framework to perform collaborative recommendation that fuses user preference, social influence, and geographical influence. Furthermore, the model uses a linear fusion framework to integrate ranked lists provided by the three recommender systems. 
        Kefalas et al. \cite{kefalas2017time} use contextual pre-filtering of the information to select the most relevant proximate users for the recommendations. The spatial influence of users' reviews represents the impact of proximate users who reviewed similar businesses to the target user. The model extends the item-based contextual filtering in two ways, (i) by leveraging the proximity factor when computing the similarity of two users and (ii) by considering the history of proximate user reviews. Liu et al. \cite{liu2013learning} introduce a geographical probabilistic factor analysis framework for POI recommendation. To learn geographical user preferences, the model encodes the spatial influence and user mobility into the user check-in process. Furthermore, the model adopts a Bayesian probabilistic non-negative latent factor model for encoding both the spatial influence and personalized preferences.
    \subsection{Semantic Effect}
        Every POI has some properties and two POIs are semantically close to each other if they consist of similar properties. Every human also has his/her own preferences and a person wants to visit those POIs which are matched to his/her preferences \cite{ye2011semantic}. So, from the check-in list of a user, we can capture the user preferences and try to predict those POIs that are semantically very similar to previous check-ins. \\
        \cite{ye2011semantic, kefalas2017time, wu2019long, li2018next, chang2018content} are some of the papers that use the semantic information from the datasets to predict the next POI. Kefalas et al. \cite{kefalas2017time} utilize the textual influence among the reviews that refer to the similarity between the reviews. Ye et al. \cite{ye2011semantic} uses a semantic annotation technique for POI networks to automatically annotate all places with category tags. The annotation algorithm learns a binary SVM classifier for each tag in the tag space to support multi-label classification. This algorithm extracts features and handles semantic annotation from places with the same tag and the relatedness among places. Wu et al. \cite{wu2019long} learns the short and long-term contextual features of POIs and leverage attention mechanisms to capture users' preference. Li et al. \cite{li2018next} propose an encoder-decoder neural network model that leverages the embedding method to incorporate heterogeneous contextual factors associated with each check-in activity, to populate the semantics of check-ins. The paper embeds check-in user and time, numerical factors, and categorical factors in contexts. Chang et al. \cite{chang2018content}  utilize the text content that provides information about the characteristics of a POI. They also measure the correlation between words by calculating the Jaccard similarity of POIs in their text content. The text content layer treats text content as a sentence and trains the word embedding vector using Word2Vec \cite{mikolov2013efficient}. All these works employed different techniques only to capture the semantic features from the POIs for getting better recommendation performance.
    \subsection{Social Influence}
        Humans are social beings. So, the decisions of a person greatly depend on his/her social status, friends, neighbors, culture, etc. These social influences affect a person's interest in visiting a POI. The probability of a user visiting a POI is increased when his/her friends give good reviews about that POI \cite{chang2018content}. To tackle the cold start problems, the social circle of a new user can be heavily beneficial and thus, models can learn a user's preferences by suggesting the same POIs of his/her social circle. \\
        \cite{christoforidis2019reline, chang2018content, kefalas2017time} use social influence to improve their POI predictions. Christoforidis et al.\cite{christoforidis2019reline} incorporated social influence alongside spatial and temporal context and combined the graphs into a unified prediction model. Kefalas et al. \cite{kefalas2017time} try to capture the social influence using users' reviews. The users having similar vocabularies are considered to be related. Here, the social influence corresponds to the correlation between the target user and others concerning the lexical analysis of their reviews. The abundance of works leveraging social influence proves the importance of social effect in POI recommendation.
    \subsection{Temporal Influence}
        Human-lives consist of fixed time patterns. A user shows distinct check-in preferences at different hours of the day and tends to have similar preferences in consecutive hours than non-consecutive hours \cite{gao2013exploring}. User activities are influenced by time. A user will go to a restaurant rather than a bar at noon and people may tend to go to visit places when a holiday appears. \\
        \cite{yuan2013time,gao2013exploring, xie2016learning,christoforidis2019reline,doan2019attentive} considered the temporal effects in POI recommendation systems. Gao et al.\cite{gao2013exploring} introduced a temporal state to represent hours of the day. Then they defined the time-dependent user check-in preferences using the temporal state. The paper proposes a temporal regularization to minimize an objective function using temporal coefficients. Their proposed framework, LRT, consists of temporal division, temporal factorization, and temporal aggregation. Yuan et al.\cite{yuan2013time} perform collaborative filtering by exploiting other user's temporal preferences to POIs. To capture the fixed routine of users' daily mobility, the model splits time into hourly slots and model the temporal preference to POIs of a user in a time slot by the POIs visited by the user in this time slot. They leverage a time factor when computing the similarity between two users and consider the historical check-ins at a time in the repository. Doan et al. \cite{doan2019attentive} uses an attention mechanism designed for utilizing spatio-temporal information. 
        
\begin{table*}[!t]
\small
\begin{threeparttable}
\caption{Descriptions of Influential Factors}
\label{influenceModel}
\centering
\begin{tabular}{ |p{80mm}|c|c|c|c|c|}
    \hline
    \multicolumn{1}{|c|}{\multirow{2}{*}{\textbf{Reference}}} & \textbf{Sequential} & \textbf{Geographical} & \textbf{Semantic} & \textbf{Social} & \textbf{Temporal} \\
    & \textbf{Effect} & \textbf{Influence} & \textbf{Effect} & \textbf{Influence} & \textbf{Influence} \\
    \hline
    STGN\cite{zhao2019stgn}, LSTPM\cite{sun2020go}, TMCA\cite{li2018next}, ASTEN\cite{doan2019attentive}, Bi-STDDP\cite{xi2019modelling}, GT-HAN\cite{liu2019geographical}, Flashback\cite{yang2020location}, TEMN\cite{zhou2019topic}, ST-RNN\cite{liu2016predicting}, HST-LSTM\cite{kong2018hst}, DeepMove\cite{feng2018deepmove}, SANST\cite{guo2020sanst} & \cmark & \cmark & \xmark & \xmark & \cmark \\ \hline
    LSPL\cite{wu2019long}, PLSPL\cite{wu2020personalized}, iMTL\cite{zhang2020interactive}, ASPPA\cite{zhao2020discovering}, ARNN\cite{guo2020attentional}, CARA\cite{manotumruksa2018contextual}, GT-HAN\cite{liu2020exploiting}, t-LocPred\cite{liu2020time}, CatDM\cite{yu2020category}, GE\cite{xie2016learning}, MGRU\cite{kala2019context} & \cmark & \cmark & \cmark & \xmark & \cmark \\ \hline
    CAPE\cite{chang2018content} & \xmark & \cmark & \cmark & \cmark & \xmark \\ \hline
    GeoSAN\cite{lian2020geography}, Geo-Teaser\cite{zhao2017geo} & \cmark & \cmark & \xmark & \xmark & \xmark \\ \hline
    CAPRE\cite{chang2020content} & \cmark & \cmark & \cmark & \cmark & \xmark \\ \hline
    STA\cite{liu2017learning} & \xmark & \cmark & \cmark & \xmark & \cmark \\ \hline
    JLGE\cite{christoforidis2018recommendation}, RELINE\cite{christoforidis2019reline}, APOIR\cite{zhou2019adversarial} & \cmark & \cmark & \xmark & \cmark & \cmark \\ \hline
    LLRec (Teacher)\cite{wang2020next} & \cmark & \cmark & \cmark & \cmark & \cmark \\ \hline
    HME\cite{feng2020hme} & \cmark & \cmark & \cmark & \xmark & \xmark \\ \hline
    DYSTAL\cite{xiong2020dynamic} & \xmark & \cmark & \cmark & \cmark & \cmark \\ \hline
    HCT\cite{zhang2020modeling} & \xmark & \cmark & \cmark & \xmark & \xmark \\ \hline
\end{tabular}
\end{threeparttable}
\end{table*}


\section{Shortcomings and Challenges}

\label{section:shortcomings}
    \subsection{Data Sparsity}
        Data sparsity is one of the most crucial problems of building a location recommender system because the graphs and matrices are far more sparse than most other recommender systems. From a user's point of view, a person travels to very few locations in his/her lifetime compared to the sheer huge number of POIs available to be visited. 
        Furthermore, exploring different POIs cost significantly higher than the exploration of different options in other fields of recommendation, which further exacerbates the data sparsity issue. Consequently, the relationships between users and POIs described in the datasets are genuinely sparse. Thus, it is always a challenge to design an effective POI recommendation system with this sparse dataset.
    \subsection{Cold Start}
        When a user joins an LBSN network, the lack of proper characterization of that user results in poor initial recommendation performance. Similarly, when a new POI is created to be explored, it heavily lags behind the already existing POIs in terms of recommendation due to the lack of historical trajectories associated with that POI. Since this problem is common to most of the present recommendation models, eliminating the cold start problem is a promising research direction in POI recommendation.
        
    \subsection{Scarcity of benchmark dataset}
        While popular check-in datasets like Foursquare, Gowalla, Yelp, Weeplaces are largely used in the previous works, the sheer scale of these datasets makes them impossible to work with all at once. Consequently, most of the prior works take only a subset (e.g. check-ins of a city within a specific timeframe) of these datasets to evaluate the proposed models. Furthermore, recent advancements in POI recommendation show that besides check-ins, data from other modalities such as textual description, user reviews, etc. can improve the recommendation performance. However, popular datasets do not contain these data which results in the proposal of new datasets (e.g. Instagram dataset) that contain these data. The variety of datasets used in different models makes it very difficult to compare the performance against different state-of-the-art models. A benchmark dataset containing data from various modalities can resolve these issues creating a solid ground to assess the performance across different models.
    \subsection{Lack of online learning}
        Analyzing different models in this review, we see that most of the POI recommendation models in the literature use offline learning i.e. models can be trained only once with the available data before deployment. In a real-world scenario, this strategy is not optimal, since everyday users are generating tons of new check-in data which carries crucial information about changes in user preferences. An online learning strategy that can update the models as new data gets available is thus of significant importance so that the model can provide optimal recommendation performance over time even in changing circumstances.
        
    \subsection{Privacy Preserved POI Recommendation}
        Like many other location-based services, user privacy is a major bottleneck for the proliferation of POI recommendation systems. Users are not willing to share their GPS traces in many cases as from location traces adversaries can reveal many sensitive and private information of the user. Thus there is an increasing need to devise a privacy preserved POI recommendation system. A couple of non-deep learning approaches, Liu et al. \cite{liu2017privacy} and Chen et al. \cite{chen2020privacy} preserve the privacy of the user data. However, since these approaches also require aggregating data from users in a central location, there are still lots of privacy concern exist. We envision that POI recommendation techniques can exploit a new domain of privacy-preserving learning, namely federated learning, that does not require accumulating user data in any central site.
       
        Essentially, federated learning is a machine learning technique that trains an algorithm across multiple devices using their local data samples, without exchanging them. In traditional distributed learning all data are distributed across multiple centralized servers which do not ensure the privacy of user data and data security. But in federated learning, a user does not need to share their data. They can train a model using their local data and share the model parameters. So, federated learning ensures data privacy and data sparsity issues. 
        In this context, Wang et al. \cite{wang2020next} recently proposed a model LLRec by generating teacher and student models. While the idea is close to the notion of federated learning, it does not fully take advantage of the parameter sharing like federated learning. Thus, developing federated learning-friendly models is an interesting research direction in this domain.

    \subsection{Recommendation for Social Groups}
        Previous works mostly focus on personalized POI recommendations where user's historical check-ins as well as other attributes are taken into account. However, POI recommendation for a group of users has mostly been out of focus in the literature. Recommendation for a social group is significantly different from personalized POI recommendation, since each group member may have different preferences for choosing POIs. The social aspect becomes particularly important when recommending POIs for a group of users, which most of the present personalized POI recommendation models are unable to handle properly. Wang et al. \cite{wang2020group} used matrix factorization and clustering techniques for group POI recommendation. But a simple model like this fails to utilize external features which largely limits the model performance. Consequently, a deep learning-based POI recommendation model for social groups can be particularly helpful for group tourism.

\section{Conclusion}

In this paper, we have presented a comprehensive survey on deep learning based POI recommendation systems. We have presented insightful findings of a plethora of recent research papers in this emerging area of research. We have categorized the POI recommendation models based on different deep learning paradigms and compare their competitive (dis)advantages. We have also presented the performance results of these techniques w.r.t. different performance metrics for different real datasets. We have identified different factors that impact the POI recommendations and provided a tabular analysis of each factor. Finally, we have discussed a series of future works on POI recommendation that provide a guideline for new researchers in this domain. To the best of our knowledge, this work is the first comprehensive review of deep learning based POI recommendations.

.

\balance
\bibliographystyle{elsarticle-num} 
\bibliography{MyCollection}

\end{document}